\documentclass[11pt,a4paper]{article}

\usepackage[truedimen,margin=30mm]{geometry} 

\usepackage{mathrsfs}
\usepackage{amssymb}
\usepackage{amsmath}
\usepackage{ascmac}
\usepackage{amsthm}
\usepackage[dvipdfmx]{graphicx}
\usepackage{natbib}
\usepackage{setspace}

\usepackage{color}
\usepackage{charter} 

\usepackage{comment} 
\usepackage{bm} 

\usepackage{subcaption} 

\usepackage{bbm}
\usepackage{url}

\usepackage{ulem}

\usepackage{titlesec}
\titleformat*{\section}{\large\bfseries}
\titleformat*{\subsection}{\it}

\def\ep{{\varepsilon}}
\def\si{{\sigma}}
\def\om{{\omega}}

\def\Si{{\Sigma}}

\def\A{{\text{\boldmath $A$}}}
\def\B{{\text{\boldmath $B$}}}
\def\C{{\text{\boldmath $C$}}}
\def\h{{\text{\boldmath $h$}}}
\def\I{{\text{\boldmath $I$}}}
 
\def\u{{\text{\boldmath $u$}}}
\def\v{{\text{\boldmath $v$}}}
 
\def\y{{\text{\boldmath $y$}}}

\def\bmu{{\text{\boldmath $\mu$}}}
 
\def\bpi{{\text{\boldmath $\pi$}}}
\def\etab{{\text{\boldmath $\eta$}}}

\def\bSi{{\text{\boldmath $\Si$}}}
 
\def\bPhi{{\text{\boldmath $\Phi$}}}
\def\bTh{{\text{\boldmath $\Theta$}}}

\def\bep{{\text{\boldmath $\ep$}}}

\def\one{{\text{\boldmath $1$}}}
\def\zero{{\text{\boldmath $0$}}}

\title{{\bf State-Space Modeling of Shape-constrained Functional Time Series}\footnote{\today}}

\date{}

\begin{document}

\maketitle
\doublespacing

\vspace{-1.5cm}
\begin{center}
{\large Daichi Hiraki$^1$, Yasuyuki Hamura$^2$, Kaoru Irie$^3$ and Shonosuke Sugasawa$^4$}

\medskip
$^1$Graduate School of Economics, The University of Tokyo, \\
$^2$Graduate School of Economics, Kyoto University \\
$^3$Faculty of Economics, The University of Tokyo \\
$^4$Faculty of Economics, Keio University
\end{center}

\vspace{1cm}
\begin{center}
{\bf \large Abstract}
\end{center}

\vspace{-0cm}
Functional time series data frequently appears in econometric analyses, where the functions of interest are subject to some shape constraints, including monotonicity and convexity, as typical of the estimation of the Lorenz curve. This paper proposes a state-space model for time-varying functions to extract trends and serial dependence from functional time series while imposing the shape constraints on the estimated functions. The function of interest is modeled by a convex combination of selected basis functions to satisfy the shape constraints, where the time-varying convex weights on simplex follow the dynamic multi-logit models. To enable posterior computation by an efficient Markov chain Monte Carlo method, a novel data augmentation technique is devised for the complicated likelihood of this model. The proposed method is applied to the estimation of time-varying Lorenz curves, and its utility is illustrated through numerical experiments and analysis of panel data of household incomes in Japan.


\newpage
\section{Introduction}

Functional data has been seen in many scientific fields, including economics and social sciences \citep[e.g.][]{horvath2012inference, kokoszka2017introduction}, where the object to be analyzed is expressed in the form of functions. Such a function is often limited to some subclass of functions, or its functional shape is constrained by monotonicity, convexity and other functional properties. Naturally, the estimation of functions under multiple constraints have been of research interest in statistics and econometrics. 

Typically, such shape constraints can be addressed in terms of the differences of functional values. 
Modeling the differences of the functional values directly leads to the realization of shape-constrained functions. 
For example, a univariate function that varies gradually is modeled by specifying the prior distribution of the first-order difference between two consecutive functional values (e.g., see \citealt{faulkner2018locally} and references therein). 
The use of a truncated prior can easily realize the monotonically increasing/decreasing functions on discrete support \citep{okano2023locally}. 
This approach can also define stochastic processes on functional spaces, including the square of Gaussian processes for derivatives of functions \citep{wang2016estimating,lenk2017bayesian,kobayashi2021flexible}, the Gaussian process truncated on constrained functional space \citep{ray2020efficient} and spline regression models with constrained coefficients \citep{shively2009bayesian,shively2011nonparametric}. 
However, the posterior inference of the aforementioned approaches requires complicated Markov chain Monte Carlo (MCMC) methods, which hinders the extension to the hierarchical and dynamic models for multiple functions. 

From the viewpoint of functional time series, several models for time-varying functions have been studied. Examples include autoregressive processes \citep{king2019bayesian} and shrinkage processes for trend estimation \citep{kowal2017bayesian,wakayama2024functional}, but they do not address shape constraints.
An exception is the Bayesian non-parametric approach to monotonic functions in \cite{canale2016bayesian}, which combines MCMC and approximation Bayesian computation for posterior inference. 

In this research, we consider a flexible modeling of time-varying monotone and/or convex functions. 
Our model is based on convex combinations of basis functions that satisfy monotonicity and/or convexity with time-varying weights constrained on the simplex. 
In modeling the dynamic convex weights, we simply take the inverse softmax transformation of the weights to define the real-valued state vector, then model its dynamics by the standard, Gaussian autoregressive process. This approach results in the state space models with the non-linear observational equation, for which the posterior sampling algorithm is not trivial. For posterior computation, we propose a novel data augmentation approach and prove that the model of interest is conditionally a dynamic linear model, enabling fast sampling of state variables by filtering and smoothing. The key idea of our augmentation is to compute the exponentiated quadratic term of the weight vector and write it as the binomial likelihood, to which the well-known P\'{o}lya-gamma augmentation can be applied \citep{polson2013bayesian,glynn2019bayesian}.

As an important application of the shape-constrained functional inference, we focus on the problem of estimating the Lorenz curves based on the aggregated income data. The Lorenz curves, together with the Gini coefficients computed from the curves, have long been utilized in economics as the measures of economic inequalities. By definition, the Lorenz curve must be monotonically non-decreasing and convex, hence the inferential problem of the Lorenz curve falls well within the scope of our study on the inference for the shape-constraint functions. 
The standard approach to inference on the Lorentz curve is to fit a hypothetical income distribution and derive the Lorenz curves from the estimated income distribution. 
The models and methods of parameter estimation related to this approach include generalized methods of moments \citep{hajargasht2012inference}, mimimum distance estimation \citep{hajargasht2020minimum}, Dirichlet likelihood \citep{chotikapanich2002estimating} and approximate Bayesian computation \citep{kobayashi2019approximate}. 
Among them, the (generalized) Dirichlet model has been extended to the state state models for the time-varying Lorentz curve \citep{kobayashi2022bayesian}. 
However, the use of Dirichlet likelihood could lead to severe misspecification of the observation model, in addition to the need to specify a particular parametric class of income distributions. 
Our approach requires no distributional assumption on income distributions, hence can define a flexible, nonparametric model for the Lorenz curve.

The rest of the paper is organized as follows. 
In Section~\ref{sec:SSM}, we introduce our proposed state-space model and its application to the time-varying Lorenz curve. 
The detailed posterior computation algorithm is discussed in Section~\ref{sec:pos}.
We demonstrate the numerical performance of the proposed method through simulation studies in Section~\ref{sec:sim}, and application to Japanese income data in Section~\ref{sec:app}.
Concluding remarks are given in Section~\ref{sec:conc}.
Additional computational details and numerical results are provided in the Supplementary Material.

{\bf Notation.} \ 
For $r$-dimensional vector $\bm{x}$,  $\bm{x}_{-i}$ is the $(r-1)$-dimensional sub-vector obtained by deleting $i$-th entry of $\bm{x}$ for $i=1,\dots, r$. For $r \times c$ matrix $\bm{X}$, $i=1,\dots ,r$, and $j=1,\dots ,c$, we denote the $(i,j)$ entry of $\bm{X}$ by $(\bm{X})_{i,j}$, and the $i$-th row vector of $\bm{X}$ with $j$-th column entry being deleted by $(\bm{X})_{i,-j}$. $(\bm{X})_{-i,j}$ is defined similarly. 
$(\bm{X})_{-i,-j}$ is the $(r-1)\times(c-1)$ submatrix of $\bm{X}$ with the $i$-th row and $j$-th column being deleted. 
$\I_r$ is the identity matrix of size $r$. $\one_r = (1,\dots, 1)^{\top}$ is the $r$-dimensional vector of ones, while $\mathbbm{1}[\cdot]$ means the indicator function. 
$\mathrm{N}_r(\bmu, \bSi)$ is the $r$-dimensional multivariate normal distribution with mean $\bmu$ and variance $\bSi$, and we write $\mathrm{N} = \mathrm{N}_1$. We also write its density function evaluated at $\bm{x}$ by $\mathrm{N}_r(\bm{x}; \bmu, \bSi)$. $\mathrm{TN}_{(a,b)}(\mu, \sigma ^2)$ is the truncated normal distribution supported on interval $(a,b)$. $\mathrm{IG}(a,b)$ is the inverse-gamma distribution with shape $a>0$ and rate $b>0$. $\mathrm{U}(a,b)$ is the uniform distribution over $(a,b)$.

\section{State-Space Models}\label{sec:SSM}

\subsection{State-space models using basis functions}
Suppose that we are interested in time-varying function $f_t(x)$ on $x\in \mathcal{X}$ for $t=1,\ldots,T$. 
We do not observe $f_t(x)$ directly, but $\widehat{f}_t(x)$ for some pre-specified augment values of $x$, which are contaminated by additive noises as 
\begin{equation*}
\widehat{f}_t(x_k) = f_t(x_k) + \epsilon _{tk}, \ \ \ \ \ \ k=1,\dots, K, 
\end{equation*}
where the distribution of noise $\epsilon _{tk}$ is defined later in (\ref{eq:SSM}). 
In what follows, we assume that the number of observed points $K$, as well as the points themselves $(x_1,\dots, x_K)$, does not change over time, but our method can be easily generalized to the situation where $K$ and $(x_1,\dots, x_K)$ can vary across time points.

Given the shape constraints of interest (e.g., monotonicity and/or convexity), denote the subclass of functions under the constraints by $\mathcal{F}$. We further assume that $\mathcal{F}$ is closed under convex combination; this condition is satisfied if $\mathcal{F}$ is the set of monotone and/or convex functions. 
In our application, $\mathcal{F}$ is the set of all possible Lorenz curves, or increasing and convex functions on $\mathcal{X}=[0,1]$. 
We model the unknown function as a convex combination of $L$ basis functions in $\mathcal{F}$:   
$$
f_t(x)= \sum_{\ell=1}^L \pi_{t\ell}h(x; a_\ell, b_\ell ), 
$$
where $h(\cdot ;a_\ell,b_\ell) \in \mathcal{F}$ for $\ell = 1,\dots ,L$, $(a_\ell,b_{\ell})$ is a pair of fixed parameters that define the basis function, and $(\pi _{t1},\dots , \pi _{tL})$ is on the $(L-1)$-dimensional simplex for all $t$ in the sense that 
$$
\sum_{\ell=1}^L\pi_{t\ell}=1 \ \ \ \ \ \mathrm{and} \ \ \ \ \ \ \pi _{t\ell} \ge 0 \ \ \ \mathrm{for} \ \ \ell =1,\dots, L.
$$ 
Since $h(\cdot ;a_\ell,b_\ell) \in \mathcal{F}$ and $\mathcal{F}$ is closed under convex combination, we have $f_t(\cdot ) \in \mathcal{F}$, or the shape constraints are always imposed on $f_t(\cdot)$. 
The choice of the number of basis functions $L$, as well as the basis functions $h(\cdot ;a_\ell,b_\ell)$, is essential in flexibly modeling of the target function. Examples of the basis functions are given in Section~\ref{sec:Lorenz} and the effect of the choice of the basis functions on inference is investigated in details in Sections~\ref{sec:sim} and \ref{sec:app} by numerical studies. 
Using a sufficiently large $L$ and appropriate basis functions, most of the functions in $\mathcal{F}$ are expected to be approximated accurately by the convex combination above. Theoretical studies that support this approach include \cite{mallick1994generalized}, where the beta cumulative distribution functions are shown to be dense in the class of continuous distribution functions on $[0,1]$.

It is also noteworthy that, in our model, the basis function $h(\cdot ;a_\ell,b_\ell)$ does not change over time, but the convex weight vector $\pi_{t\ell}$ does. Hence, the model for $\pi_{t\ell}$ characterizes the dynamics of function $f_t(\cdot)$. 
To introduce a time-series structure in the convex weights, we first define real-valued state variable $u_{t\ell} \in \mathbb{R}$ by mapping weight vector $\pi_{t\ell} \in (0,1)$ by the inverse softmax function below:
$$
\pi_{t1}=\frac{1}{1+\sum_{\ell'=1}^{L-1}\exp(u_{t\ell'})}
, \ \ \ \ \ \ 
\pi_{t,\ell+1}=\frac{\exp(u_{t\ell})}{1+\sum_{\ell'=1}^{L-1}\exp(u_{t\ell'})}, \ \ \ \ \ell=1,\ldots,L-1.
$$
Then, for $(L-1)$-dimensional vector $\u_t=(u_{t1},\ldots,u_{t,L-1})$, we introduce a vector autoregressive model as the state equation, namely, 
\begin{equation}\label{eq:VAR} 
 \u_t = (\I_{L-1}-\bPhi)\bmu + \bPhi \u_{t-1} + \bep_t, \ \ \ \ \bep_t\sim \mathrm{N}(0, \bSi),  
\end{equation}
where $\bmu$ is the mean vector, and $\bPhi$ and $\bSi$ are the unknown coefficient and covariance matrix of size $L-1$, respectively. In our application, we assume that $\bPhi = \mathrm{diag}(\phi _1,\dots, \phi _{L-1})$ and $\bSi = \mathrm{diag}(\sigma^2 _1,\dots, \sigma^2 _{L-1})$, where $\phi _l \in (-1,1)$ and $\sigma _l^2 >0$ for $\ell=1,\dots, L-1$, hence each $u_{t \ell}$ follows a univariate AR(1) process independently. Priors for $(\bmu,\bPhi,\bSi)$ and initial value $\u_0$ are set so that the AR(1) processes become stationary, as explained later in Section~\ref{sec:gibbs}.

Let $y_{tk}=\widehat{f}_t(x_k)$ and $h_{k\ell}=h(x_k;a_\ell,b_\ell)$. 
The observational equation for $y_{tk}$ is set as 
\begin{equation}\label{eq:SSM}
y_{tk}|\u_{t}\sim \mathrm{N}\left(\sum_{\ell=1}^L\pi_{t\ell}(\u_t)h_{k\ell}, \nu_{tk}^2\right), \ \ \ \ k=1,\ldots,K. 
\end{equation}
The state equation (\ref{eq:VAR}) and observational equation (\ref{eq:SSM}) define the state space model we propose. 
The choice of models for observational variance $\nu _{tk}^2$ is open to the end users. In our study, we set $\nu_{tk}^2 = \nu^2$ for all $t$ and $k$ and define a prior distribution for $\nu^2$; the observational noises are independent and identically distributed across time $t$ and argument $x_k$. The prior for $\nu^2$ is given in Section~\ref{sec:gibbs}.

Since we observe the functional value with the Gaussian additive noise, the mean of $y_{tk}$ under the model (\ref{eq:SSM}) is $f_t(x_k)$. In other words, we assume that the mean function is constrained in $\mathcal{F}$. To be precise, for augment $x_k$, we have 
\begin{equation} \label{eq:mean}
    \mathrm{E}[ y_{tk} | f_t(x_k), \nu _{tk}^2 ] = f_t(x_k) = \sum_{\ell=1}^L\pi_{t\ell}(\u_t)h(x_k;a_\ell,b_\ell). 
\end{equation}
The variance of $y_{tk}$ equals the error variance $\nu_{tk}^2$. 
This shows that the proposed state space model is indeed the time series extension of the additive noise model, where the error variance does not contain the information about the mean function.

\subsection{Difference from mixture approach}
\label{sec:mixture}

Another standard approach to the constrained mean function is to use a mixture of normals, 
where the means of the mixture components are the basis functions. 
We here address why such standard approach is not preferable in the context of our research, despite its computational simplicity.  
If one models response $y_{tk}$ not by the convex combination of the basis functions, but by the mixture of them, the model is expressed as 
\begin{equation*}
y_{tk}|\u_{t}\sim \sum _{\ell=1}^L \pi_{t\ell}(\u_t) \mathrm{N}\left( h_{k\ell}, \nu_{tk}^2\right), \ \ \ \ k=1,\ldots,K. 
\end{equation*}
Both the mixture model and ours have the same mean in (\ref{eq:mean}), thus we can achieve the shape constraints on the mean function by either model. However, our model in (\ref{eq:SSM}) has the variance $\mathrm{Var}[ y_{tk} |\u_t ] = \nu_{tk}^2$, while the variance of the mixture model is complicated as 
\begin{equation*}
\mathrm{Var}[ y_{tk} |\u_t ] = \nu_{tk}^2 + \sum_{\ell=1}^L\pi_{t\ell}(\u_t)h_{k\ell}^2 - \left( \sum_{\ell=1}^L\pi_{t\ell}(\u_t)h_{k\ell} \right) ^2.
\end{equation*}
Clearly, our model separates the observational variance from the mean structure, while the mixture model has the variance to depend on the mean. This separation in the usage of parameters helps the interpretation and prior elicitation of those parameters. 

Note also that, in the mixture model, using the same variance parameter of $\nu _{tk}^2$ across the mixture components is restrictive in the sense that all the components have the same variance. Thus, in using the mixture model, we should use a variance parameter customized for each mixture component, or $\nu _{tk\ell}^2$, instead of $\nu _{tk}^2$. In the numerical examples, where we set $\nu_{tk}^2 = \nu ^2$ in our model, it is fair to use the following mixture model: 
\begin{equation} \label{eq:mixture}
y_{tk}|\u_{t}\sim \sum _{\ell=1}^L \pi_{t\ell}(\u_t) \mathrm{N}\left( h_{k\ell}, \nu_{\ell }^2\right), \ \ \ \ k=1,\ldots,K, 
\end{equation}
where we use independent inverse gamma prior for $\nu_{\ell}^2$. The posterior inference for this mixture model is straightforward, for we can use the P\'{o}lya-gamma augmentation directly. For details on the prior settings and computation, see the Supplementary Materials.

\subsection{Example: Estimation of time-varying Lorenz curve} \label{sec:Lorenz}

As an example of application of the proposed model, we extensively study the estimation of the time-varying Lorenz curve. Here, we do not specify the income distribution explicitly, but work directly on the Lorenz curve under the required shape constraints. We denote the Lorenz curve at time $t$ by $f_t(\cdot)$. The Lorenz curve maps income level, $x \in [0,1]$, to the share of the total income, $f_t(x) \in [0,1]$. Lorenz curve $f_t(\cdot)$ must be monotonically non-decreasing, and satisfy the boundary conditions, $f_t(0)=0$ and $f_t(1)=1$ (e.g.,~\citealt{rasche1980functional}). 
In addition, if the income distribution is continuous, then the Lorenz curve must be convex. 
Since we model the Lorenz curve as the convex combination of the basis functions, we must impose the shape constraints and boundary conditions listed above onto the basis functions. 

We consider two classes of the basis functions. One is the cumulative distribution functions of the beta distribution \citep{gelfand1998model}; 
\begin{equation*}
    h(x; a_\ell, b_\ell) = \int _0^x \frac{1}{B(a_\ell, b_\ell)} t^{a_\ell-1}(1-t)^{b_\ell-1} dt, 
\end{equation*}
where $a_\ell>0$ and $b_\ell>0$. We call this choice the beta basis function. Since this basis function is the cumulative distribution function of the continuous distribution supported on $[0,1]$, the required shape constraints and boundary conditions are satisfied. 
The value of the basis function, or the incomplete beta function, can be numerically evaluated. 

Another class of basis functions we consider is given by $h(x; a_\ell, b_\ell)  = (1-(1-x)^{a_\ell })^{1/b_\ell}$, where $a_\ell ,b_\ell \in (0,1]$ \citep[e.g.][]{rasche1980functional}. 
For convenience, we call this form the Pareto basis function, for this class includes the Lorenz curve of the Pareto distribution of income when $a_\ell < 1$ and $b_\ell = 1$. 
It is easy to verify that this basis function also satisfies the shape constraints and boundary conditions. The Gini coefficient of this basis function, as explained below, is obtained in a simple form. 

The Gini coefficient (or Gini index) summarizes the Lorent curve into a single numerical value as the measure of economic inequality. Given Lorenz curve $f_t(\cdot )$, the Gini coefficient is defined by 
\begin{equation*}
    G_t = 1 - 2 \int _0^1 f_t(x) dx,
\end{equation*}
or the twice of the area between the Lorenz curve and 45-degree line. For most income distributions and Lorenz curves, this integral is intractable and requires intensive computational efforts for posterior inference, such as sequential Monte Carlo methods \citep{kobayashi2019approximate}. 
However, since we have expressed the Lorenz curve as the convex combination of the basis functions, we can simplify the expression of the Gini coefficient above as 
\begin{equation*}
    G_t = \sum _{k=1}^K \pi _{t\ell} G_{\ell}, 
\end{equation*}
where $G_{\ell}$ is the Gini coefficient of the $\ell$-th basis function given by 
\begin{equation*}
    G_{\ell} = 1 - 2 \int _0^1 h (x;a_{\ell},b_{\ell}) dx. 
\end{equation*}
Thus, the computation of the Gini coefficient reduces to that of $G_{\ell}$, which can be evaluated easily for both the beta and Pareto basis functions. In particular, the Gini coefficient of the Pareto basis function is available in the closed form, or $G_{\ell} = 1-2\mathrm{B}(1/a_{\ell},1/b_{\ell} + 1) / a_{\ell}$ \citep{rasche1980functional}. 
Given the values of $G_\ell$'s, the Gini coefficient is the function of convex weights $\pi _{t\ell}$. 
Once the posterior samples of $\pi _{t\ell}$ are obtained, then we can construct the samples of $G_t$ for posterior inference. This is convenient especially in computing the posterior quantiles to assess the uncertainty about $G_t$, as well as the posterior mean/median as point estimates.

\section{Posterior computation}\label{sec:pos}
\subsection{Overview}
In this section, we derive an augmented model for posterior inference under the proposed model and provide the Gibbs sampler algorithm. Given the observed data $\y=(\y_1^\top,\ldots,\y_T^\top)$, the posterior distribution of $\u=(\u_1^\top,\ldots,\u_T^\top)$ and $\bTh=\{ \bmu , \bPhi, \bSi\}$ is given by  
\begin{align*}
\pi(\u, \bTh | \y)\propto \pi(\Theta) \prod_{t=1}^T \prod_{k=1}^{K}\mathrm{N}\bigg(y_{tk}; \sum_{\ell = 1}^{L} h_{k\ell}\pi_{t \ell }( \u_t ), \nu_{tk}^2\bigg)  \mathrm{N}_{L-1}(\u_t;(\I_{L-1}-\bPhi)\bmu + \bPhi\u_{t-1}, \bSi).
\end{align*}
In implementing the Gibbs sampler, one must sample from the conditional posteriors of $\bTh$ and $\u$. For parameters $\bTh$, the conjugate priors are available and utilized for posterior sampling. For $\u_t$, the conditional posterior distribution does not have a familiar form since $\pi_{t \ell }( \u_t )$ is a nonlinear function of $\u_t$. 

To obtain an efficient sampling scheme for $\u_t$, we propose a novel data augmentation technique using Poisson and P\'{o}lya-gamma random variables. 
Since $\pi_{t \ell }( \u_t )$ is the softmax function of $\u_t$, one might expect that the P\'{o}lya-gamma augmentation would become applicable as practiced in the Bayeian analysis of the multinomial regression models \citep{glynn2019bayesian}. 
Unfortunately, this is not the case in our model, because the likelihood of $\u_t$ is the {\it exponentiated} binomial likelihood, where $\pi_{t \ell }( \u_t )$ appears in the augment of the exponential function. To this likelihood, the P\'{o}lya-gamma augmentation cannot be directly applied. 
Below, we show that additional Poisson-distributed latent variables make the likelihood of $\u_t$ be conditionally binomial, which enables the P\'{o}lya-gamma augmentation.

\subsection{Augmentation by latent variables}
For notational simplicity, we write $\bpi_t(\u_t)=(\pi_{t1}(\u_t),\ldots,\pi_{tL}(\u_t))^\top$ and $\h _k = ( h_{k 1} , \dots , h_{k L} )^{\top }$. 
First, noting that $\one_L^{\top}\bpi_t(\u_t) = 1$, we have 
\begin{align*}
\prod_{k=1}^{K}\mathrm{N}\bigg(y_{tk}; \sum_{\ell = 1}^{L} h_{k\ell}\pi_{t \ell }(\u_t), \nu_{tk}^2\bigg) 
\propto  \exp \Big\{-{\bpi_t}(\u_t)^{\top } \A_t \bpi_t(\u_t)\Big\},
\end{align*}
where $\A_t=\sum_{k=1}^{K}(y_{tk}\one_L - \h _k ) (y_{tk}\one_L - \h _k )^\top / (2 \nu_{tk}^2 )$.  
Next, we transform the latent $(L-1)$-dimensional vector $u_{t\ell}$ in the exponential scale by setting $v_{t \ell }=\exp(u_{t\ell})$ for $\ell=1,\ldots,L-1$ and define $L$-dimensional vector $\v_t=(1,v_{t1},\ldots,v_{t,L-1})^\top$. Then, it holds that ${\bpi_t}(\u_t) = (\one_L^\top \v_t)^{-1}\v_t$ and ${\bpi_t}(\u_t)^{\top } \A_t \bpi_t(\u_t)=(\one_L^\top \v_t)^{-2}\v_t^\top \A_t \v_t$.

We propose the following augmentation for each $v_{t \ell }$. 
Let $\v_{t,-\ell}$ be an $(L-1)$-dimensional vector obtained by deleting $v_{t \ell }$ in $\v_t$. Define $s_{t\ell}=\one_{L-1}^\top \v_{t,-\ell}$, which does not contain $v_{t \ell }$. 
Then, we have
\begin{align*}
\exp \Big\{-(\one_L^\top \v_t)^{-2}\v_t^\top \A_t \v_t\Big\}
\propto 
\exp \bigg\{ - \frac{1}{(v_{t \ell } + s_{t\ell})^2} (v_{t \ell },s_{t\ell})\B_{t \ell } (v_{t \ell },s_{t\ell})^\top
\bigg\},
\end{align*}
where $\B_{t \ell }$ is a symmetric $2\times 2$-matrix with entries $b_{t\ell}$, $c_{t\ell}$ and $d_{t\ell}$ and defined as 
\begin{align*}
\B_{t \ell } = \begin{pmatrix} b_{t \ell } & c_{t \ell } \\ c_{t \ell } & d_{t \ell } \end{pmatrix} 
= 
\begin{pmatrix} 1 & 0 \\ \zero & \v_{t,-\ell}/s_{t\ell} \end{pmatrix}^{\top} \begin{pmatrix} ( \A _t )_{\ell + 1, \ell + 1} & ( \A _t )_{\ell + 1, - (\ell + 1)} \\ ( \A _t )_{- (\ell + 1), \ell + 1} & ( \A _t )_{- (\ell + 1), - (\ell + 1)} \end{pmatrix} \begin{pmatrix} 1 & 0 \\ \zero & \v_{t,-\ell}/s_{t\ell} \end{pmatrix}.
\end{align*}
Computing this quadratic form further, we obtain 
\begin{align*}
\exp &\Big\{-{\bpi_t}(\u_t)^{\top } \A_t \bpi_t(\u_t)\Big\} \\
&\propto 
\exp \bigg\{ \max ( b_{t \ell }, d_{t \ell } )- \frac{1}{(v_{t \ell } + s_{t\ell})^2} 
(v_{t \ell },s_{t\ell})\B_{t \ell } (v_{t \ell },s_{t\ell})^\top
\bigg\}  \\
&= 
\exp \bigg[ \frac{1}{(v_{t \ell } + s_{t\ell})^2} 
(v_{t \ell },s_{t\ell})
\Big\{\max(b_{t \ell }, d_{t \ell })\one_2\one_2^{\top}-\B_{t \ell }\Big\} 
(v_{t \ell },s_{t\ell})^\top
\bigg]  \\
&= \exp \bigg[ {|b_{t \ell } - d_{t \ell }|\{v_{t \ell }^2 \mathbbm{1}(b_{t \ell } < d_{r \ell }) + s_{t\ell}^2 \mathbbm{1}(b_{t \ell } > d_{t \ell })\} + 2 \big\{ \max(b_{t \ell },d_{t \ell }) - c_{t \ell }\big\} v_{t \ell } s_{t\ell} \over (v_{t \ell }+s_{t\ell})^2} \bigg].
\end{align*}
The final formula above is the product of two exponential functions and has the following series expression:
\begin{align*}
\sum_{z_{t\ell 1} = 0}^{\infty } \sum_{z_{t\ell 2} = 0}^{\infty } 
\frac{|b_{t \ell } - d_{t \ell }|^{z_{t\ell 1}}}{z_{t\ell 1}!} 
\frac{[2 \{ \max(b_{t \ell }, d_{t \ell }) - c_{t \ell }\}]^{z_{t\ell 2}}}{z_{t\ell 2}!} 
\frac{v_{t \ell }^{2 z_{t\ell 1} \mathbbm{1}(b_{t \ell } < d_{t \ell }) + z_{t\ell 2}}}{(v_{t \ell } + s_{t\ell})^{2 (z_{t\ell 1} + z_{t\ell 2})}} 
s_{t\ell}^{2 z_{t\ell 1} \mathbbm{1}(b_{t \ell } > d_{t \ell }) + z_{t\ell 2}},
\end{align*}
which is viewed as the mixture using Poisson-distributed latent variables $z_{t\ell 1}$ and $z_{t\ell 2}$.
Noting that $v_{t \ell }=\exp(u_{t\ell})$, the binomial likelihood of $u_{t\ell}$ is seen in the expression above, to which the P\'{o}lya-gamma augmentation is applied as 
\begin{align*}
\frac{v_{t \ell }^{2 z_{t\ell 1} \mathbbm{1}(b_{t \ell } < d_{t \ell }) + z_{t\ell 2}}}{(v_{t \ell } + s_{t\ell})^{2( z_{t\ell 1} + z_{t\ell 2})}} 
& \propto \exp\Big[ u_{t\ell}\big\{2 z_{t\ell 1} \mathbbm{1}(b_{t \ell } < d_{t \ell }) + z_{t\ell 2} - ( z_{t\ell 1} + z_{t\ell 2})\big\}\Big] \\
&\quad \times \int_{0}^{\infty } 
\exp \Big\{ - {\om_{t\ell} \over 2} ( u_{t\ell} - \log s_{t\ell} )^2 \Big\}
f_{\rm PG}(\om_{t\ell}; 2(z_{t\ell 1} + z_{t\ell 2}), 0)d\om_{t\ell},
\end{align*}
where $f_{\rm PG}(\cdot; b, c)$ is the density function of ${\rm{PG}}(b,c)$, or the P\'{o}lya-gamma distribution with parameters $b>0$ and $c\in\mathbb{R}$.
In this expression, we can read off the linear and Gaussian likelihood of $u_{t\ell}$; we will discuss how this expression is utilized in computing the full conditional of $u_{t\ell}$ in the next subsection. 

The full conditional distributions of the newly-introduced latent variables can be read-off easily.
The full conditionals of $z_{t\ell 1}$ and $z_{t\ell 2}$ are mutually independent and 
\begin{equation}\label{eq:z-full}
\begin{split}
&z_{t\ell 1} \sim {\rm{Po}} \Big( {|b_{t \ell } - d_{t \ell }| [ v_{t \ell }^2 \mathbbm{1}(b_{t \ell } < d_{t \ell }) + s_{t\ell}^2 \mathbbm{1}(b_{t \ell } > d_{t \ell })] \over (v_{t \ell } + s_{t\ell} )^2} \Big) \\
&z_{t\ell 2} \sim {\rm{Po}} \Big( {2 \{ \max(b_{t \ell }, d_{t \ell }) - c_{t \ell }\} v_{t \ell } s_{t\ell} \over (v_{t \ell } + s_{t\ell} )^2} \Big),
\end{split}
\end{equation}
where $\rm{Po}(\lambda)$ is the Poisson distribution with mean $\lambda > 0$. 
Furthermore, the full conditional of $\om_{1 \ell },\ldots,\om_{T \ell }$ are mutually independent and the conditional distribution of $\omega_{t \ell }$ is ${\rm{PG}} (2 (z_{t\ell 1} + z_{t\ell 2} ), u_{t\ell} - \log s_{t\ell})$.

\subsection{Sampling of latent $\u_t$}

For each $t$, conditional on $z_{t\ell 1}$, $z_{t\ell 2}$, $\om_{t\ell}$ and $\u_{t,-\ell}$, the likelihood of $u_{t\ell}$ becomes 
\begin{equation*}
    \exp \Big\{ u_{t\ell} z_{t\ell 1} \{ 2  \mathbbm{1}(b_{t \ell } < d_{t \ell })  -  1 \} - {\om_{t\ell} \over 2} ( u_{t\ell} - \log s_{t\ell} )^2 \Big\},
\end{equation*}
being proportional to $\mathrm{N}( \tilde{y}_{t\ell}; u_{t\ell},1/\om_{t\ell})$ with fictitious data $\tilde{y}_{t\ell} = \log s_{t\ell} + z_{t\ell 1}\{ 2  \mathbbm{1}(b_{t \ell } < d_{t \ell })  -  1 \} / \om_{t\ell}$. This conditional likelihood is Gaussian and linear in $u_{t\ell}$ and convenient for posterior computation together with the Gaussian prior defined in (\ref{eq:VAR}). 
However, note that we have conditioned $\u_{t,-\ell}$ to obtain this expression, hence the full conditional we are working on here is not the joint distribution of $\u_t$, but the conditional distribution of $u_{t\ell}$ for fixed $\ell$. Accordingly, the conditional prior must be computed.

For some $1<t<T$ and $\ell$, consider the full conditional posterior of $u_{t\ell}$. The prior used in this computation consists of $p(\u_t |\u_{t-1},\bTh )$ and $p(\u_{t+1}|\u_t,\bTh )$, both of which are defined from the expression in (\ref{eq:VAR}). The former is further computed as the function of $u_{t\ell}$ and proportional to the conditional distribution of $p(u_{t\ell} | \u_{t,-\ell},\u_{t-1},\bTh)$, and the latter is also obtained similarly.  
While these are the univariate normal distributions, their conditional means and variances involve the manipulation of matrices of size $L-1$. In general, computing the conditional means and variances at every MCMC iteration could be cumbersome and computationally costly. 

In our application, we set $\bPhi = \mathrm{diag} (\phi _1,\dots, \phi_{L-1})$ and $\bSi=\mathrm{diag}(\sigma_1^2,\dots ,\sigma_{L-1}^2)$ as often assumed in econometric analysis, so that the prior in (\ref{eq:VAR}) becomes the independent, univariate AR processes with stationary mean $\bmu = (\mu _1,\dots ,\mu _{L-1})^{\top}$. This simplifies the expression of $p(\u_t |\u_{t-1},\bTh )$ in terms of $u_{t\ell}$ as $\mathrm{N}(u_{t\ell}; (1-\phi_\ell ) \mu _\ell + \phi_{\ell} u_{t-1,\ell},\sigma_{\ell}^2)$. Furthermore, since the prior process of $\{ u_{t\ell} \} _{t=0,1,\dots ,T}$ is independent of the others $\{ u_{t\ell'} \} _{t=0,1,\dots ,T}$ for $\ell'\not= \ell$, the full conditional posterior of $\{ u_{t\ell} \} _{t=0,1,\dots ,T}$ becomes a multivariate normal distribution and can be obtained as the joint posterior of the pseudo model:
\begin{equation} \label{eq:dlm}
\begin{split}
 \tilde{y}_{t\ell} | u_{t\ell} &\sim \mathrm{N}( u_{t\ell} , 1/\om_{t\ell} ), \ \ \ \ \ \ \tilde{y}_{t\ell} = \log s_{t\ell} + \frac{ z_{t\ell 1} \{ 2  \mathbbm{1}(b_{t \ell } < d_{t \ell })  -  1 \} }{ \om_{t\ell} }, \\
    u_{t\ell} | u_{t-1,\ell} &\sim \mathrm{N}( (1-\phi_\ell)\mu_\ell +\phi_l u_{t-1,\ell} , \sigma _{\ell }^2 ).
    \end{split}
\end{equation}
This is a dynamic linear model, for which an efficient algorithm of simulation from the conditional posterior is known as the simulation smoother \citep{de1991diffuse,de1995simulation}. The simulation of state variables under this pseudo DLM is integrated into posterior computation as one step of the Gibbs sampler algorithm \citep{carter1994gibbs,fruhwirth1994data}.

\subsection{Summary of Gibbs sampler} \label{sec:gibbs}

To complete the Gibbs sampler algorithm, we detail the sampling step for $\bTh = (\bmu, \bPhi ,\bSi)$ and $\nu ^2$. In general, under the conjugate priors for $(\bmu, \bPhi ,\bSi )$, the full conditionals become normal, truncated (univariate) normal and inverse-Wishart distributions, respectively, from which it is easy to simulate. In our application, where we assume $\bPhi = \mathrm{diag} (\phi_1,\dots, \phi _{L-1})$ and $\bSi = \mathrm{diag} (\sigma^2 _1,\dots, \sigma^2_{L-1})$, we use independent conjugate priors, $\mu_\ell \sim \mathrm{N}( \overline{m}_{0\ell}, \overline{v}_{0\ell}^2) $, $\phi_\ell \sim \mathrm{TN}_{(-1,1)}( m_{0\ell}, v_{0\ell}^2)$, $\sigma_{\ell}^{2}\sim \mathrm{IG}(  n_{0\ell} /2, d_{0\ell} /2 )$ for $\ell=1,\ldots,L-1$ and $\nu^{2} \sim \mathrm{IG}(n_0/2, d_0/2)$. 
We also assume the stationarity of $\{ \u_t \}$ and employ the prior $u_{0\ell}|\mu_{\ell},\phi_{\ell}, \sigma_{\ell}^2 \sim \mathrm{N}(\mu_\ell ,\sigma_{\ell}^2/(1-\phi_{\ell}^2))$.

The sampling steps for $u_{t\ell}$'s, the unknown parameters and the latent variables are summarized as follows: 
\begin{itemize}

\item Sampling of parameters; Conditional on $\u_t$, and with the other latent variables being marginalized out, generate the samples of parameters as follows. 

\item[-] 
{\bf (Sampling of $\bPhi = \mathrm{diag} (\phi _1,\dots, \phi _{L-1})$)} \ For $\ell=1,\ldots,L-1$, denote the sample obtained at the previous iteration by $\phi_\ell^{\rm old}$. Then, we generate a candidate $\phi_\ell^{\rm new} \sim \mathrm{TN}_{(-1,1)}( m_{1\ell}, v_{1\ell}^2)$, where
\begin{equation*}
\begin{split}
&m_{1\ell} = v_{1\ell}^2 \left(\sigma_\ell^{-2} \sum_{t=1}^T (u_{t,\ell}-\mu_\ell)(u_{t-1,\ell}-\mu_\ell) + v_{0\ell}^{-2} m_0\right), \\
&v_{1\ell}^2 = \left(\sigma_\ell^{-2} \sum_{t=2}^T (u_{t-1,\ell}-\mu_\ell)^2 + v_{0\ell}^{-2}\right)^{-1},
\end{split}
\end{equation*}
and accept it with probability
\begin{equation*}
\min \left\{ 1, \sqrt{\frac{1-(\phi_\ell^{\rm new})^2}{1-(\phi_\ell^{\rm old} )^2}} \right\}.
\end{equation*}

\item[-] 
{\bf (Sampling of $\bSi = \mathrm{diag} (\sigma^2 _1,\dots, \sigma^2_{L-1})$)} \ For $\ell=1,\ldots,L-1$, the full conditional distribution of $\sigma^2_{\ell }$ is $\mathrm{IG}(  n_{1\ell} /2, d_{1\ell} /2 )$, where 
\begin{equation*}
\begin{split}
n_{1\ell} = T + n_{0\ell} + 1, \quad 
d_{1\ell} = \sum_{t=1}^{T} (u_{t,\ell} - \phi_{\ell } u_{t-1, \ell})^2 + d_{0\ell} + (1 - \phi_{\ell}^2)u_{0,\ell}^2.
\end{split}
\end{equation*}

\item[-] 
{\bf (Sampling of $\bmu = (\mu_1,\dots, \mu_{L-1})^{\top}$)} \ For $\ell=1,\ldots,L-1$, the full conditional distribution of $\mu_\ell$ is $\mathrm{N}( \overline{m}_{1\ell}, \overline{v}_{1\ell}^2)$, where
\begin{equation*}
\begin{split}
&\overline{m}_{1\ell} = \overline{v}_{1\ell}^2 \left( \frac{1-\phi_\ell}{\sigma_\ell^2}\sum_{t=1}^T (u_{t, \ell } - \phi_\ell u_{t-1,\ell}) + \frac{(1-\phi_\ell^2)u_{0,\ell}}{\sigma_\ell^2} + \frac{\overline{m}_{0\ell}}{\overline{v}_{0\ell}^2} \right), \\
&\overline{v}_{1\ell}^2 = \left( {1-\phi_\ell^2 + T(1-\phi_\ell)^2 \over \sigma_\ell^2} + \overline{v}_{0\ell}^{-2} \right)^{-1}.
\end{split}
\end{equation*}

\item[-]
{\bf (Sampling of $\nu^2$)} \ The full conditional distribution of $\nu^2$ is $\mathrm{IG}(n_1/2, d_1/2)$, where 
\begin{equation*}
n_1 = \sum_{t=1}^T K + n_0, \quad
d_1 = \sum_{t=1}^{T} \sum_{k=1}^{K} \Big[ \sum_{\ell=1}^{L} \pi _{t \ell } ( \u _t ) ( y_{t k} - h_{k\ell} ) \Big]^2 + d_0.
\end{equation*}

\item Sampling of state variables; For each $\ell = 1,\dots ,L-1$, generate $\{ (z_{t\ell 1},z_{t\ell 2},\omega _{t\ell} ) \} _{t=1,\dots,T}$ and $ \{ u_{t\ell} \} _{t=0,1,\dots, T}$, as follows.  

\item[-] 
{\bf (Sampling of latent $z_{t\ell 1}$ and $z_{t\ell 2}$)}\ \ Generate $z_{t\ell 1}$ and $z_{t\ell 2}$ from (\ref{eq:z-full}). 

\item[-]
{\bf (Sampling of latent $\om_{t\ell}$)}\ \
Generate $\om_{t\ell}$ from ${\rm{PG}} (2 (z_{t\ell 1} + z_{t\ell 2} ), u_{t\ell} - \log s_{t\ell})$.

\item[-]
{\bf (Sampling of $u_{t\ell}$)}\ \ Generate $\{ u_{t,\ell} \} _{t=0,\dots ,T}$ by using the simulation smoother to the pseudo dynamic linear models in (\ref{eq:dlm}).

\end{itemize}

Owing to the novel data augmentation, all the sampling steps are simply the direct simulation from familiar distributions. 
Although the proposed algorithm introduces multiple latent variables and needs iterative updates of each element of $\u_t$, the mixing of the Markov chains is quite reasonable in our numerical examples, as we confirm in Section~\ref{sec:sim}.

\section{Simulation Studies}\label{sec:sim}

In this section, together with the real data analysis of Section~\ref{sec:app}, we investigate the performance of the proposed model and computational method numerically. The goal of our numerical study is threefold: the illustration of the proposed method, the sensitivity analysis of the choice of basis functions, and the comparison with other possible approaches. 

\subsection{Simulation data}

Throughout this section, we use the dataset simulated from the process described in this subsection. We set $T=200$ and consider two choices of the true number of basis functions: $K\in\{4,9\}$. For $k=1,\dots,K$, the augments are set as $x_k = 0.2k$ ($K=4$) or $x_k = 0.1k$ ($K=9$). The basis functions used to generate the synthetic data are the $L=3$ Beta basis functions with parameters $(a_\ell,b_\ell)=(1, 1)$, $(3.0, 1)$, and $(1, 0.3)$. Then, we generate the synthetic data, $\{ y_{tk} \} _{t=1,\dots,T}$ for each $k$, from the additive noise model below: 
\begin{align*}
&y_{tk}|u_{t}\sim \mathrm{N}\left(\sum_{\ell=1}^L\pi_{t\ell}(u_t)h_{k\ell}, \nu^2\right), \ \ \ \ k=1,\ldots,K,\\
&u_{t\ell}|u_{t-1,\ell}\sim \mathrm{N}((1-\phi_\ell)\mu_\ell + \phi_{\ell}u_{t-1,\ell}, \sigma^2_{\ell}),  \ \ \ \ \   \ell=1,2, \\
&u_{0\ell} \sim \mathrm{N}( \mu_\ell, \sigma _\ell ^2 / (1-\phi _\ell^2) ), \ \ \ \ \   \ell=1,2,
\end{align*}
where we set $(\mu _1,\mu_2) = (0.1,-0.3)$, $\nu^2=0.01^2$ and $\sigma_1^2 = \sigma _2^2 =0.005$. For coefficient $(\phi_1,\phi_2)$, we set $\phi _1 = \phi _2$ and consider three cases: $\phi _\ell \in \{ 0.9, 0.95, 0.99 \}$.

\subsection{Settings and illustration of the proposed models}

For the simulated dataset, we apply the proposed state space models, defined by (\ref{eq:VAR}) and (\ref{eq:SSM}), with diagonal $\bPhi$ and $\Si$. 
For the unknown parameters, we employ prior distributions, $\nu^2 \sim \mathrm{IG}(0.001/2, 0.001/2)$, $\mu _\ell \sim \mathrm{N}(0,5^2)$, $\phi _{\ell} \sim \mathrm{TN}_{(-1,1)}(0.8, 0.2^2)$, and $\sigma^2_{\ell} \sim \mathrm{IG}(0.001/2, 0.001/2)$ for $\ell = 1,2$.

We first confirm that the posterior inference under the proposed model is successful when using the same basis functions as those in the data generating process. 
In this experiment, we generated 30,000 posterior samples after discarding initial 10,000 MCMC samples as burn-in periods. The raw computational time to generate those samples in the case of $K=9$ was 240 minutes when the proposed Gibbs sampler was executed in \textsf{R} on our laptop computer with 1.6GHz Dual-Core Intel Core i5 processor equipped with 8 GB of RAM. 
The posterior means are close to the true values, and the 95\% credible intervals cover most of the true values. 
The effective sample sizes (ESS) are sufficiently large in all scenarios, showing the efficiency of the proposed Gibbs sampler. For more details, see the Supplementary Materials.

\subsection{Comparison with mixture approach }

We next consider the mixture model--- an easier approach to the shape constraint on mean functions--- and observe its data fit in the additive noise situation. The mixture model we consider has the observational equation in (\ref{eq:mixture}), while the same state equation (\ref{eq:VAR}) is used for the convex weights. The prior for component variance is set as $\nu _\ell ^2 \sim \mathrm{IG}(0.01/2, 0.01/2)$ for all $\ell$ independently. 
The basis functions, or the component means of the mixture, are those used in the data generating process. The prior for the other parameters, as well as the number of the MCMC iterations and burn-in period, are the same.

The comparison with the proposed method is made via the posterior distribution of convex weights $\pi_{t\ell}$, Gini coefficient $G_t$, and $f_t(x_k)$ for each time $t$ and $k$ computed by the MCMC samples. Note that the Gini coefficient under the mixture model (\ref{eq:mixture}) is the weighted average of the Gini coefficients of the basis functions and easily evaluated. 
Table~\ref{tab:sim_95_pi_Gini} presents the root mean squared errors (RMSE) of posterior means, empirical coverage probabilities (CP) and average lengths (AL) of 95\% credible intervals for $\pi_{t\ell}$ and $G_t$ averaged over $t = 1,...,T$ under the proposed functional state space models (FSSM) and mixture models. 
The CPs of the FSSM are around the nominal level for all the parameters, showing reasonable posterior uncertainty quantification. The RMSE and AL values of the FSSM under $K=9$ are smaller than those under $K=4$ since more data are observed when $K=9$. 
By contrast, the CPs of the mixture approach are all significantly lower than the nominal level. This undercoverage, together with the higher RMSEs, indicates a misfit of the mixture model in the additive noise situation. 
A possible reason for this is that, as pointed out in Section~\ref{sec:mixture}, parameter $\pi_{t\ell}$ is used in modeling not only the mean function but also the observational variance, hence the inference on the convex weights is strongly affected by the observational noises.

\begin{table}[htbp]
    \centering
    \begin{tabular}{cccccccccc}
        \hline
         &  &  & $K = 4$ &  & & $K = 9$ &   \\
         &  & RMSE & AL & CP & RMSE & AL & CP \\
         \hline
        $\phi_{\ell} = 0.90$ & $\pi_{t\ell}$ & 2.009 & 0.075 & 0.947 & 1.477 & 0.054 & 0.943 \\
        (FSSM) & $G_t$ & 0.558 & 0.020 & 0.935 & 0.428 & 0.016 & 0.930 \\
        \hline
        $\phi_{\ell} = 0.95$ & $\pi_{t\ell}$ & 1.828 & 0.072 & 0.960 & 1.441 & 0.053 & 0.940 \\
        (FSSM) & $G_t$ & 0.500 & 0.020 & 0.955 & 0.421 & 0.016 & 0.925 \\
        \hline
        $\phi_{\ell} = 0.99$ & $\pi_{t\ell}$ & 1.715 & 0.069 & 0.967 & 1.377 & 0.052 & 0.945 \\
        (FSSM) & $G_t$ & 0.489 & 0.019 & 0.955 & 0.417 & 0.016 & 0.930 \\
        \hline
        $\phi_{\ell} = 0.90$ & $\pi_{t\ell}$ & 8.267 & 0.121 & 0.288 & 8.354 & 0.090 & 0.178 \\
        (Mixture) & $G_t$ & 2.678 & 0.054 & 0.555 & 2.599 & 0.041 & 0.400 \\
        \hline
        $\phi_{\ell} = 0.95$ & $\pi_{t\ell}$ & 9.785 & 0.113 & 0.252 & 9.606 & 0.089 & 0.170 \\
        (Mixture) & $G_t$ & 3.694 & 0.053 & 0.355 & 3.572 & 0.043 & 0.270 \\
        \hline
        $\phi_{\ell} = 0.99$ & $\pi_{t\ell}$ & 13.113 & 0.123 & 0.230 & 13.297 & 0.096 & 0.165 \\
        (Mixture) & $G_t$ & 6.329 & 0.056 & 0.160 & 6.262 & 0.045 & 0.100 \\
         \hline
    \end{tabular}
    \caption{The root mean squared errors multiplied by 100 (RMSE), average lengths of 95\% credible intervals (AL) and coverage probabilities (CP) of the credible intervals for $\pi_{t\ell}$ and $G_t$ averaged over $t = 1, ..., T$ in the case of known basis functions for $3\times 2 = 6$ datasets.}
    \label{tab:sim_95_pi_Gini}
\end{table}

\subsection{Other models and basis functions}

We next investigate the estimation performance of the proposed and other methods, using the synthetic datasets of $\phi _{\ell}=0.95$. 

\begin{itemize}
\item[-]
{\bf The proposed model (FSSM)}:\ So far, the basis functions of the data generating process have been used in estimation as well (referred to as ``oracle'').  
Here, we also implement a set of seven Pareto basis functions (referred to as ``misspecified'') with parameters $(a_\ell,b_\ell)$ = $(1, 1)$, $(0.7, 0.6)$, $(0.9, 0.25)$, $(0.8, 1)$, $(0.25, 0.9)$, $(0.9, 0.5)$, and $(0.6, 1)$. The same prior and MCMC settings are used for this model.

\item[-]
{\bf Autoregressive Gaussian process model (ARGP)}:\ Gaussian processes are the standard models for functional data. Here, we use a Gaussian process to model the evolution of the time-varying function. The model has the state space representation as: 
\begin{align*}
\y_t &= \boldsymbol{\theta}_t + \bep_t,  \ \ \ \  \bep_t\sim \mathrm{N}(\zero, \sigma^2 \I_K),
\\
    \boldsymbol{\theta}_{t+1} &= \phi \boldsymbol{\theta}_t + \etab_t, \ \ \ \ \etab_t\sim \mathrm{N}(\zero, \C),
\end{align*}
where $\boldsymbol{\theta}_t = (f_t(x_1),\dots, f_t(x_K))^{\top}$, and $(\phi , \sigma^2)$ are unknown parameters, and $\C$ is a $K\times K$ covariance matrix whose $(k,k')$-element is $\tau^2 \exp \{ -\psi |x_k-x_{k'}| \}$, or the exponential correlation function. 
Note that the shape constraints are not imposed on the estimated functions. While using this ``vanilla'' version of ARGP, we also consider the projection method \citep{lin2014bayesian} as the ``projected" ARGP--- a post-processing approach that imposes the monotonicity on the sampled functional values under the ARGP model. In implementing this model, we set $\psi^{-1} \sim \mathrm{U}(0,100)$, or the uniform distribution on interval $(0,100)$, $\tau^2 \sim \mathrm{IG}(0.01/2, 0.01/2)$, $\phi \sim \mathrm{U}(-1,1)$ and $\si ^2 \sim \mathrm{IG}(0.01/2, 0.01/2)$. 
For details on the computational method, see Chapter~11, \cite{banerjee2014hierarchical}.

\item[-]
{\bf Multivariate dynamic linear model (DLM)}:\ 
Viewing the observed functional values as the multivariate time series data, we apply a dynamic linear model. Specifically, a local level model with the random walk state evaluation is considered: 
\begin{align*}
   \y_t &= \boldsymbol{\theta}_t + \bep_t, \quad
    \bep_t \sim \mathrm{N}(0, \bm{V}_t), \\
    \boldsymbol{\theta}_{t+1} &= \boldsymbol{\theta}_{t} + \etab_t \quad
    \etab_t \sim \mathrm{N}(0, \bm{W}_t).
\end{align*}
In specifying observational and state variances, we consider two models. One is the constant variances, $\bm{V}_t = \bm{V}$ and $\bm{W}_t = \bm{W}$ with inverse-Wishart (IW) priors: $\bm{V}\sim\mathrm{IW}(5, 5 \times 0.02^2 \I_K )$ and $\bm{W}\sim\mathrm{IW}(5, 5 \times 0.01^2 \I_K)$. The other is a conjugate stochastic volatility (SV) model; we set $\bm{W}_t = w_t \bm{V}_t$, specify $w_t$ by discounting with a discount factor of $0.90$, and apply the matrix-beta inverse-Wishart processes for $\bm{V}_t$ with discount factor $0.95$. For details on the model and computational method, see Chapter~10, \cite{PradoFerreiraWest2021}.
\end{itemize}

The ARGPs and DLMs considered here are easily estimated by the standard Gibbs sampler. However, one needs more efforts for the computation of the Gini coefficients under those models, which is detailed in the Supplementary Materials.

\begin{figure}
\centering
\includegraphics[width=\linewidth]{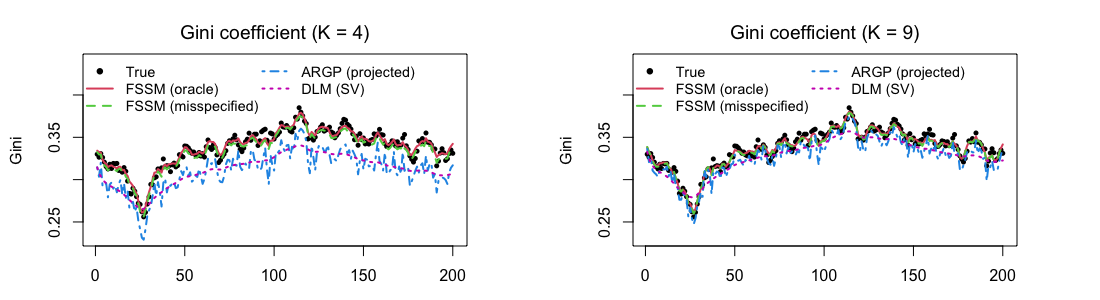}
\caption{Time series of the true and posterior means of Gini coefficients obtained by FSSM using ``oracle" and ``misspecified" basis functions, ARGP (projected) and DLM with SV.}
\label{fig:sim_gini}
\end{figure}

We first computed the posterior means of the time-varying Gini coefficients under the listed models, which are summarized in Figure~\ref{fig:sim_gini}. 
It is noteworthy that the FSSM with the misspecified Pareto basis functions can provide accurate point estimates, implying the robustness of the proposed methods to the choice of basis functions. The ARGP and DLM significantly underestimate the true Gini coefficients. The estimation accuracy improves in all models as more data becomes available ($K=9$), but the estimates closest to the true values are still provided by the FFSMs.

In addition, we conduct the posterior predictive analysis of $y_{kt}$ via the posterior predictive loss \citep{gelfand1998model} to compare the data fit of the candidate models. Specifically, the posterior predictive distributions are summarized into two measures: the sum of posterior predictive variances (PPV) and the total posterior predictive squared errors (PPSE, or the sum of variance and squares of bias). 
The results are presented in Table~\ref{tab:sim_LPPL}. 
The PPVs and PPSEs of the misspecified FSSM are only slightly larger than those of the oracle model in all scenarios, by which we confirm the successful posterior analysis even with the misspecified basis functions. The ARGP models are outperformed by the proposed methods in both measures, and the effect of post-projection is negligible in this example. The DLM with the constant variances (indicated by IW) has the smaller PPV than the FSSMs when much data are available ($K=9$), but its PPSE is larger than those of the FSSMs due to the bias caused by the limited flexibility of this DLM. 

In the Supplementary Materials, we present the RMSEs, CPs and ALs of functional values $f_t(x)$ for several augments $x$, showing the superiority of the proposed models.

\begin{table}[htb!]
\centering
\begin{tabular}{cc|cccccc}
\hline
&& \multicolumn{2}{c}{FSSM} & \multicolumn{2}{c}{ARGP} & \multicolumn{2}{c}{DLM}\\
&  & oracle & misspecified & vanilla & projected & IW & SV \\
\hline
PPV & $K = 4$ & -2.430 & -2.350 & -1.340 & -1.340 & -2.127 & -1.922 \\
 & $K = 9$ & -1.638 & -1.588 & -1.189 & -1.189 & -1.840 & -1.462 \\
PPSE & $K = 4$ & -1.861 & -1.780 & -1.072 & -1.070 & -1.280 & -1.035 \\
 & $K = 9$ & -1.022 & -0.974 & -0.839 & -0.838 & -0.959 & -0.600 \\
\hline
\end{tabular}
\caption{The sum of posterior predictive variances (PPV) and total posterior predictive squared errors (PPSE) for six methods applied to simulated data.}
\label{tab:sim_LPPL}
\end{table}

\section{Application to Japanese Income Survey Data}
\label{sec:app}

In this section, we estimate the dynamic Lorenz curves and Gini coefficients using the monthly income share data in Japan. 
The data is retrieved from the Family Income and Expenditure Survey prepared by the Ministry of Internal Affairs and Communications of Japan (available at \url{https://www.e-stat.go.jp/en}).
Our dataset contains the income shares of the $K = 4$ income classes of the 10,000 working households surveyed between January 2000 and September 2018 ($T = 225$), being adjusted to the population size. 
These classes are equally sized; each class covers 20\% of the households, hence $x_1=0.2$, $x_2 = 0.4$, $x_3=0.6$ and $x_4=0.8$. 
A similar dataset is analyzed in \cite{kobayashi2021flexible} by using the parametric Dirichlet likelihood to model time-varying Lorenz curve. 

\begin{figure}[htb!]
\centering
\includegraphics[width=0.48\linewidth]{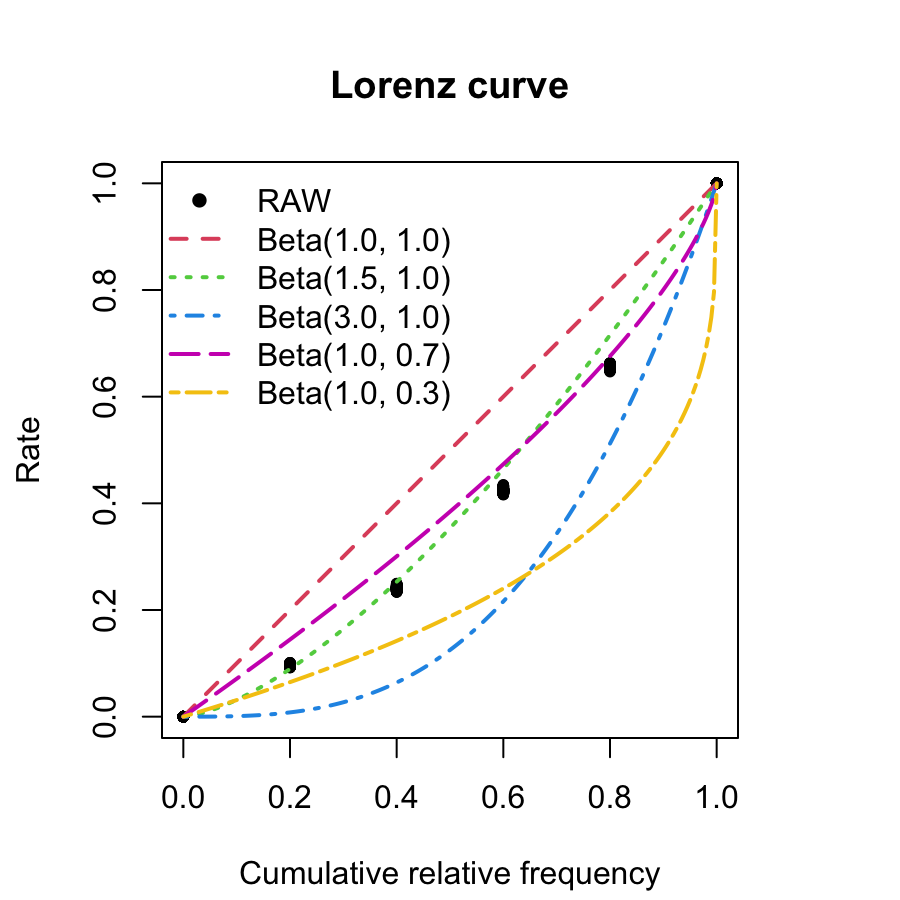}
\includegraphics[width=0.48\linewidth]{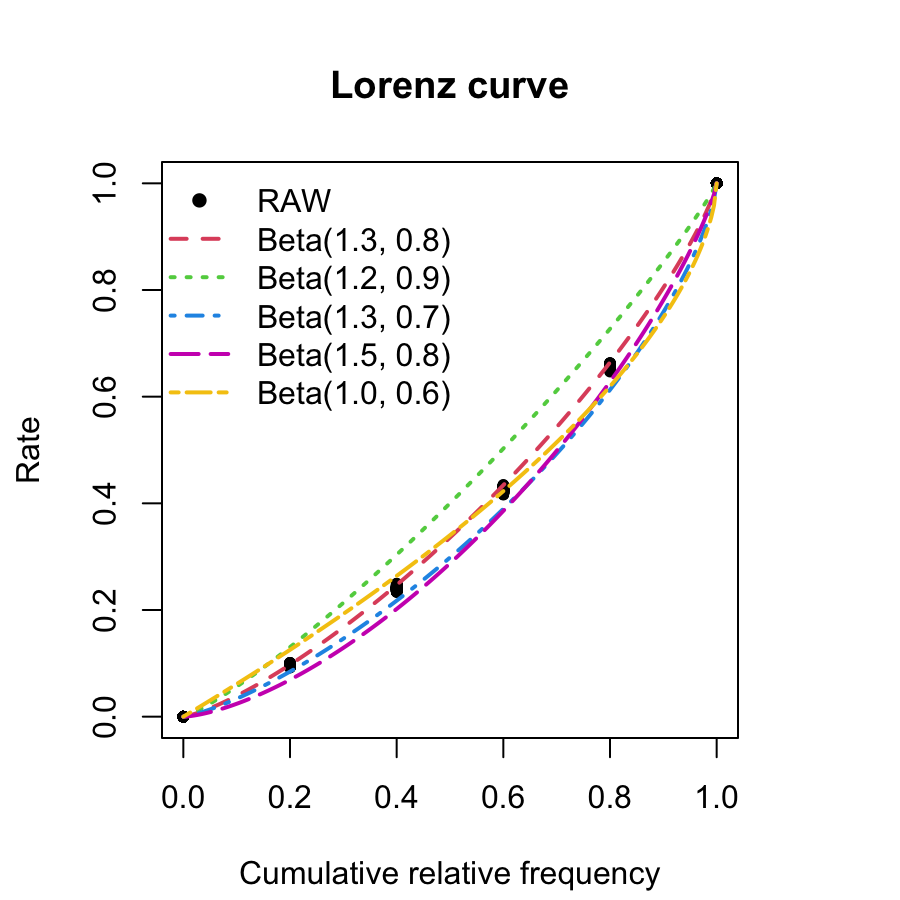}
\caption{
The observed data (black dots) and basis functions in Basis Set 1 (left) and Basis Sets 2 and 3 (right). 
}
\label{fig:ObsBasis}
\end{figure}

Here, we apply the proposed methods with beta basis functions and three sets of hyperparameters: 
\begin{align*}
{\rm {\bf Basis \ Set \ 1}}:&\ \ L=5, \ \ (a_\ell,b_\ell)=(1.0,1.0), (1.5, 1.0), (3.0,1,0), (1.0, 0.7), (1.0, 0.3),\\
{\rm {\bf Basis \ Set \ 2}}:&\ \ L=3, \ \ (a_\ell,b_\ell)=(1.2, 0.9), (1.5, 0.8), (1.0, 0.6), \\
{\rm {\bf Basis \ Set \ 3}}:&\ \ L=5, \  \ (a_\ell,b_\ell)=(1.2, 0.9), (1.5, 0.8), (1.0, 0.6), (1.3, 0.8), (1.3, 0.7). 
\end{align*}
Note that Basis Set 2 is nested in Basis Set 3. 
We plot the basis functions used in Basis Sets 1 and 3 with the observed functional values in Figure~\ref{fig:ObsBasis}. This figure confirms that these choices are not overly misspecified; the convex combination of these basis functions is expected to explain the observed functional values. 
In implementing the proposed models, we assigned the same prior distributions as those in Section~\ref{sec:sim}. 
Furthermore, we also applied the comparative methods used in Section~\ref{sec:sim}. 
In all cases, 80,000 posterior samples are generated after discarding the initial 20,000 samples as burn-in period.

First, we compute the log PPV and PPSE of each model (see Section~\ref{sec:sim} for details). The results are provided in Table~\ref{tab:real_LPPL}. 
The three proposed FSSMs have smaller PPVs and PPSEs than the ARGPs and DLMs and show their better fit to the dataset used in this example.  
Among the three basis function choices, the model with Basis Set 1 best fits the data in both measures. 
It is worth noticing that, in Figure~\ref{fig:ObsBasis}, the basis functions used in Basis Set 1 are not necessarily close to the observed functional values. 
This observation indicates that the variations in basis functions could contribute to the overall fitting.  
In the Supplementary Material, we provide the summary of posterior inference on the model parameters and time series plots of the posterior means of convex weights.

\begin{figure}[htb!]
\centering
\includegraphics[width=\linewidth]{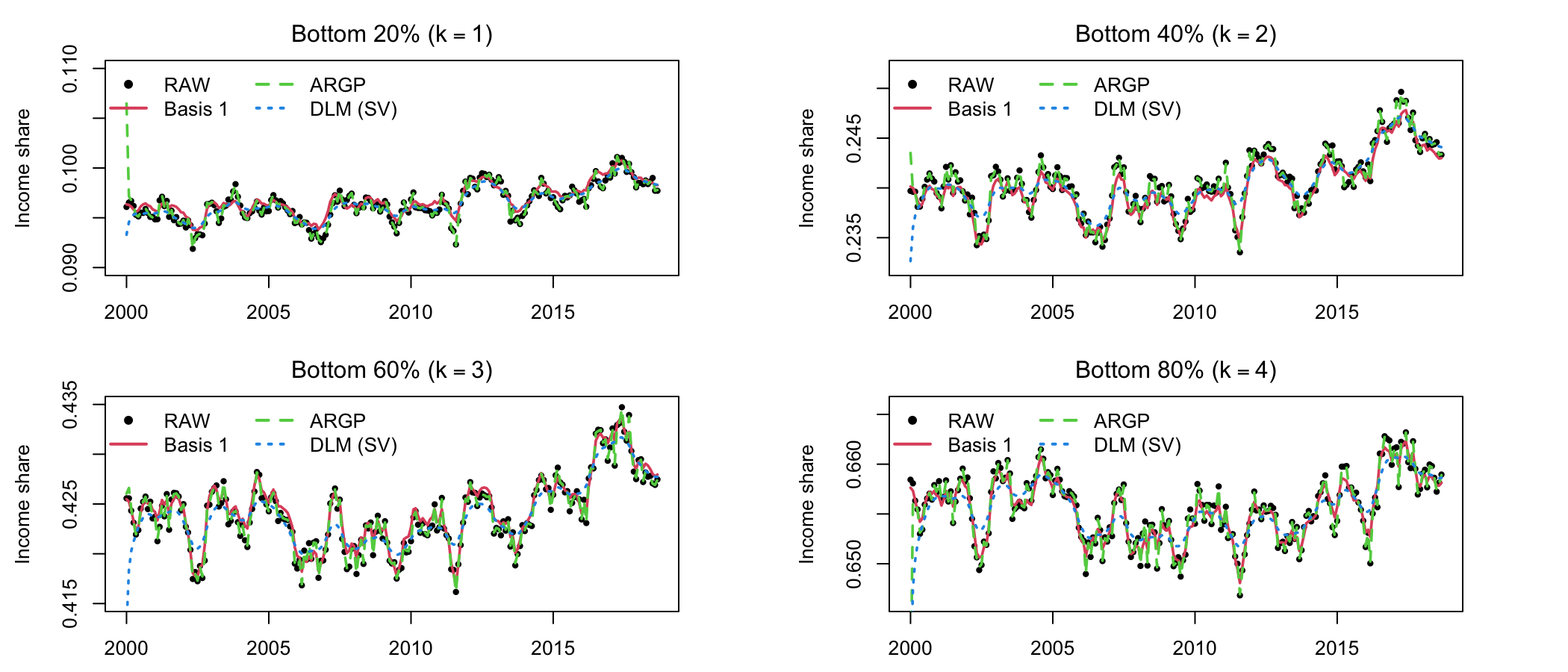}
\caption{The posterior means of the income shares for the bottom 20, 40, 60, and 80\% estimated under Basis Set 1, ARGP, and DLM (SV) models overlaid on the raw data (RAW).}
\label{fig:real_Gini1}
\end{figure}

Figure~\ref{fig:real_Gini1} presents posterior predictive means of the income shares for the bottom 20, 40, 60, and 80\% ($x=0.2,0.4,0.6,0.8$), obtained by the FSSM (Basis Set 1), ARGP and DLM.
The FSSM can successfully estimate smoothed time trends. By contrast, the DLM provides overly smoothed time trends. 
The ARGP model tends to overfit the data and fails to extract meaningful time trends, possibly due to the Gaussian process without shape restrictions being overly flexible. 

\begin{figure}[htb!]
\centering
\includegraphics[width=\linewidth]{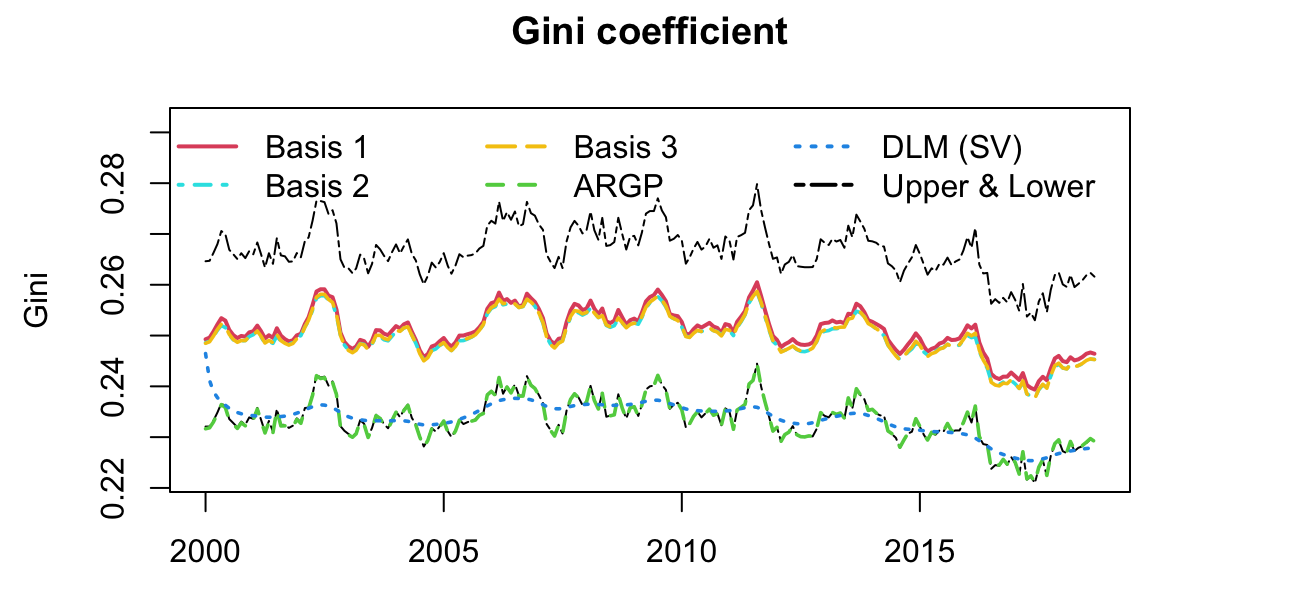}
\caption{The posterior means of Gini coefficients estimated by FSSM with Basis Sets 1, 2, and 3, ARGP, DLM and theoretical upper and lower bounds of Gini coefficients at each time.}
\label{fig:real_Gini2}
\end{figure}

In Figure~\ref{fig:real_Gini2}, we present the time series of the estimated Gini coefficients. In the same figure, we also show the non-parametric upper and lower bounds of Gini coefficients computed by using observed points \citep{mehran1975bounds}. 
The estimated Gini coefficient of the three FFSMs are almost identical and included in the intervals of the theoretical lower and upper bounds at all time points. 
By contrast, the estimates of ARGP are almost identical to the theoretical lower bound since it tends to overfit the observed data as confirmed in Figure~\ref{fig:real_Gini1}.
The DLM also underestimates the Gini coefficient; its estimates are even smaller than the theoretical lower bound.

\begin{table}[htb!]
\centering
\begin{tabular}{c|ccccccc}
\hline
& \multicolumn{3}{c}{FSSM} & \multicolumn{2}{c}{ARGP} & \multicolumn{2}{c}{DLM}\\
& Basis 1 & Basis 2 & Basis 3 & vanilla & projected & IW & SV \\
\hline
PPV & -6.009 & -5.735 & -5.718 & -2.556 & -2.556 & -2.127 & -3.720 \\
PPSE & -5.792 & -5.398 & -5.369 & -2.514 & -2.514 & \ 1.004 & -3.594 \\
\hline
\end{tabular}
\caption{The sum of posterior predictive variances (PPV) and total posterior predictive squared errors (PPSE) for seven methods applied to real data (log-scale). }
\label{tab:real_LPPL}
\end{table}

\section{Concluding Remarks}\label{sec:conc}

In this studies, we proposed a state-space modeling for time-varying functions with shape restrictions (in particular, monotone and convex functions with boundary conditions) and developed an efficient posterior computation algorithm using a novel data augmentation strategy combined with the filtering/smoothing algorithm. 
We also considered other possible approaches, including the mixture model, Gaussian processes and time series models, but they were not as competitive as the proposed model in our simulation and real data applications. 

One potential limitation of the proposed method is that it may not be able to capture abrupt structural changes in time-varying functions due to the Gaussian autoregressive models for dynamic convex weights. 
This issue could be solved by revisiting the modeling of state variables $\u_t$.  
From the viewpoint of computational statistics, it is worth emphasizing that the data augmentation strategy used in this paper can also be applied to other models involving convex weights in the location parameter. Finally, although we focused on the times series analysis in this study, the proposed state-space model could be extended to handle spatial or spatio-temporal data, which will be left to a future study.

\section*{Acknowledgement}
We thank Kazuhiko Kakamu for his helpful comments on the early version of this paper. 
This work is partially supported by Japan Society for Promotion of Science (KAKENHI) grant numbers 20H00080, 21H00699, 22K13374, 22K20132, 20J10427 and 19K11852.

\vspace{0.5cm}
\bibliographystyle{chicago}
\bibliography{ref}

\newpage
\setcounter{equation}{0}
\setcounter{section}{0}
\setcounter{table}{0}
\setcounter{figure}{0}
\setcounter{page}{1}
\renewcommand{\thesection}{S\arabic{section}}
\renewcommand{\theequation}{S\arabic{equation}}
\renewcommand{\thetable}{S\arabic{table}}
\renewcommand{\thefigure}{S\arabic{figure}}

\vspace{1cm}
\begin{center}
{\LARGE
{\bf Supplementary Material for ``State-Space Modeling of Shape-constrained Functional Time Series"}
}
\end{center}

This Supplementary Material provides the details of posterior computation algorithms of comparative methods and additional numerical results.

\section{MCMC algorithm of mixture approach}

The MCMC method for the mixture model can be derived easily. With latent variable $z_{tk}$ being introduced in the model, we have
\begin{align*}
    &y_{tk}| \u _{t} , \{ z_{tk} = \ell \} \sim N\left( h_{k\ell}, 
    \nu _{\ell}^2 \right), \\
    &\mathrm{P}[ z_{tk} = \ell |\u_t ] = \pi_{t\ell}( \u _t ), \ \ \ \ \ \ell \in \{  1,\dots,L \}.
\end{align*}
The likelihood of $\u_t$ is the softmax function:
\begin{equation*}
\mathrm{P}[ z_{t1},\ldots,z_{tK} |\u_t ] = \prod _{k=1}^{K} \pi_{t, z_{t k}} ( \u _t ) = \prod _{k=1}^{K} \left(  \frac{ e^{u_{t,z_{tk}-1}} }{ \sum _{\ell'=0}^{L - 1} e^{u_{t\ell'}} } \right),
\end{equation*}
where $\u_t$ follows the dynamic model (\ref{eq:VAR}) and $u_{t0}=0$.
The full conditional distributions are described as follows:

\begin{itemize}
\item[-] \ 
{\bf (Sampling of $z_{tk}$)} \ The full conditional probability being $z_{tk}=\ell$ is 
\begin{align*}
\frac{\phi(y_{tk}; h_{k\ell}, 
\nu _{\ell}^2 ) \exp(u_{t,\ell-1})}{\sum_{\ell'=1}^L\phi(y_{tk}; h_{k\ell'},
\nu _{\ell'}^2 ) \exp(u_{t,\ell'-1})}.
\end{align*}

\item[-] \ 
{\bf (Sampling of $u_{t\ell}$)} \ Given $u_{t\ell'}$ for $\ell'\neq \ell$, the likelihood of $u_{t\ell}$ can be augmented as 
\begin{align*}
&\frac{ \{\exp({u_{t\ell}})\}^{N_{t\ell}} }{ \left\{ \exp(u_{t\ell}) + \sum_{\ell'=0, \ell'\neq \ell}^{L-1} \exp({u_{t\ell'}}) \right\}^{K} }\\
& \ \ \ \ \ \ 
\propto \exp \left\{ \Big(N_{t\ell}-\frac{K}{2}\Big) \xi_{t\ell} \right\} \int_0^\infty \exp\left(-\frac{\omega_{t\ell} \xi_{t\ell}^2}{2}\right) p(\omega_{t\ell}|K,0) d\omega_{t\ell},
\end{align*}
where $N_{t\ell}=\sum_{k=1}^{K} \mathbbm{1}(z_{tk}=\ell+1)$, $\xi_{t\ell} = u_{t\ell} - \log\{\sum_{\ell'=0, \ell' \neq \ell}^{L-1} \exp({u_{t\ell'}})\}$ and $p(\omega_{t\ell}|K,0)$ is the density of ${\rm PG}(K, 0)$.
Note that the above augmentation can be done separately for $t=1,\ldots,T$.  
The full conditional distribution of $\omega_{t\ell}$ is ${\rm PG}(K, \xi_{t\ell})$.
Under the augmentation, the full conditional posterior of $(u_{1\ell},\ldots,u_{T\ell})$ is equivalent to the posterior of the pseudo model: 
\begin{align*}
&\tilde{y}_{t\ell} | u_{t\ell} \sim \mathrm{N}(u_{t\ell}, 1/\om _{t\ell} ), \ \ \ \ 
u_{t\ell} | u_{t-1,\ell} \sim \mathrm{N}( (1-\phi_\ell)\mu_\ell +\phi_l u_{t-1,\ell} , \sigma _{\ell }^2 ).
\end{align*}
where 
$$
\tilde{y}_{t\ell}
=\log \left\{ \sum_{\ell'=0, \ell' \neq \ell}^{L-1} \exp({u_{t\ell'}}) \right\} + \frac{2N_{t\ell} - K}{2 \om _{t\ell}} .
$$
Then, we can implement the algorithm of filtering and smoothing to generate $\{ u_{t\ell'} \} _{t=0,\dots,T}$ for $\ell'\neq \ell$.

\item[-] \ 
{\bf (Sampling of $\nu_{\ell}^2$)} \ By assuming the prior $\nu_{\ell}^{2} \sim \mathrm{IG}(n_{0\ell}/2, d_{0\ell}/2)$, the full conditional distribution is ${\rm IG}(n_{1\ell}/2, d_{1\ell}/2)$, where  
$$
n_{1\ell} =n_{0\ell} + \sum_{t=1}^T \sum_{k=1}^{K} \mathbbm{1}(z_{tk}=\ell), \ \ \ \ \ 
d_{1\ell} = d_{0\ell} + \sum_{t=1}^T \sum_{k=1}^{K} \mathbbm{1}(z_{tk}=\ell) (y_{tk} - h_{k\ell})^2.
$$
\end{itemize}

\section{The Gini coefficients under ARGPs and DLMs}

\subsection{ARGPs}

    The Gini coefficient of $f$ is given by 
    \begin{equation*}
         G = 1 - 2 \int _0^1 f(x) dx.
    \end{equation*}
    Denote the $M+1$ equally-spaced grid points on $[0,1]$ by $(z_0,\dots , z_M)$, i.e., $z_0=0$, $z_M=1$, and $z_m - z_{m-1} = 1/M$ for all $m$. Then we can approximate the integral above by the Riemann sum as 
    \begin{equation*}
         G \approx  1 - 2 M^{-1} \sum _{m=1}^M f(z_m).
    \end{equation*}
    Hence, the posterior samples of $G$ can be constructed from those of $f(z_m)$, which can be generated easily when using the ARGPs, as explained below. 

    We first note that the $K$ arguments of the observed functional values, or $\{ x_k \} _{k=1,\dots ,K}$, are not necessarily identical to the $M+1$ grid points introduced above. Here, for simplicity, we assume that $K<M+1$ and $\{ x_k \} _{k=1,\dots K}$ are included in the grid points $\{ z_m \} _{m=0,\dots ,M}$. Then, let $\boldsymbol{\theta}_t^M$ be the vector $(f_t(z_0), ..., f_t(z_M))^\top$, where $f_t(z_0) = f_t(0) = 0$ and $f_t(z_M) = f_t(1) = 1$ by the boundary conditions. Then, we can relate this vector to the mean function in the main text, $\boldsymbol{\theta}_t = ( f_t(x_1), \dots , f_t(x_K) )^{\top}$; there exists the $(K+2)\times(M+1)$ matrix $\boldsymbol{F}^M$ such that 
    \begin{equation*}
        \begin{bmatrix}
            0 \\ \boldsymbol{\theta}_t \\ 1
        \end{bmatrix} = \boldsymbol{F}^M \boldsymbol{\theta}_t^M. 
    \end{equation*}
    Similarly, we re-define the observation as the $(K+2)$-dimensional vector,
    \begin{equation*}
        \y_t^M = \begin{bmatrix}
            0 \\ \y_t \\ 1
        \end{bmatrix} = \begin{bmatrix}
            0 \\ \hat{f}_t(x_1) \\ \vdots \\ \hat{f}_t(x_K) \\ 1
        \end{bmatrix}. 
    \end{equation*}
    With these variables, the target model has the state-space representation as 
    \begin{align*}
    \y_t^M &= \boldsymbol{F}^M \boldsymbol{\theta}_t^M + \bep_t,  \ \ \ \  \bep_t\sim \mathrm{N}(\zero, \boldsymbol{V}),
    \\
        \boldsymbol{\theta}_{t+1}^M &= \boldsymbol{G}^M \boldsymbol{\theta}_t^M + \etab_t^M, \ \ \ \ \etab_t^M\sim \mathrm{N}(\zero, \C^M),
    \end{align*}
    where $\C^M$ is a $(M+1) \times (M+1)$ covariance matrix whose $(m,m')$-element is $\tau^2\exp\{ -\psi |z_m-z_{m'}| \}$ for $m, m'=2,...,M$ and the others are 0, and
    \begin{align*}
        \boldsymbol{V} = 
        \begin{bmatrix}
            0 & \dots & 0 \\
            \vdots & \si^2 \I_{K} & \vdots \\
            0 & \dots & 0
        \end{bmatrix}
        , \quad 
        \boldsymbol{G}^M =
        \begin{bmatrix}
            1 & \dots & 0 \\
            \zero & \phi \I_{M-1} & \zero \\
            0 & \dots & 1
        \end{bmatrix}.
    \end{align*}
    In Sections~\ref{sec:sim} and \ref{sec:app}, we set $M = 40$, that is, $z_m = 0.025m$ for $m=0,\dots, M$. Note that these grids include $(x_1,\dots, x_4) = (0.2,0.4,0.6,0.8)$. The priors for $(\psi, \tau^2, \phi, \sigma^2)$ are specified in the main text.

    \subsection{DLMs} 

    Unlike the ARGPs, we cannot implement the model-based interpolation of the functional values under the DLMs. That is, $f_t(x)$ is not available (or does not exist in the definition of the model) for $x\not\in \{ x_k \} _{k=1,\dots,K}$. Hence, the approximation of the Gini coefficients must rely only on the functional values on the observed points, or $\{ f_t(x_k) \} _{k=1,\dots, K}$. 
    We calculate the approximate Gini coefficients by using the posterior samples of $\boldsymbol{\theta}_t = ( \theta_{t,1},...,\theta_{t,K} )^\top$ by 
    \begin{equation*}
        G_t = 1 - \sum_{k=1}^{K+1} \frac{\theta_{t,k-1}+\theta_{t,k}}{x_k-x_{k-1}}.
    \end{equation*}
    where $\theta_{t,0}=0$, $\theta_{t,K+1}=1$, $x_0=0$, and $x_{K+1}=1$.
    The summation in the expression above is (the twice of) the area of the polygon obtained by connecting each point of $\boldsymbol{\theta}_t$. With $\boldsymbol{\theta}_t$ being replaced with observed functional values $\hat{f}_t(x_k)$, the above expression gives the non-parameteric lower bounds used in the main text.

\section{Additional Simulation Results}

\subsection{Point and interval estimates of the proposed method}
Table~\ref{tab:sim_95_mcmc} summarizes the posterior analysis under the proposed FSSM with the basis functions being correctly specified. 
The effective sample size (ESS) is confirmed to be sufficiently large in both cases of $K$. The ESS increases in $K$ due to the increase of the information at each time. 
Regarding the estimation performance, the posterior means are reasonably close to the true values, and the coverage of the true values by the $95\%$ credible intervals is successful. 
It is also reasonable that the posterior means under $K=9$ tend to be closer to the true values with shorter credible intervals than those under $K=4$ since we have more data and information with larger $K$.

\begin{table}[htbp]
    \centering
    \begin{tabular}{cc|ccc|ccc}
        \hline
         &  &  & $K = 4$ &  &  & $K = 9$ &  \\
        Parameter & True & Mean & 95\% CI & ESS & Mean & 95\% CI & ESS \\
        \hline
        $\phi_{1}$ & 0.95 & 0.959 & ( 0.910,  0.996) & 3479 & 0.960 & ( 0.914,  0.995) & 5431 \\
        $\phi_{2}$ & 0.95 & 0.936 & ( 0.840,  0.995) & 801 & 0.951 & ( 0.895,  0.994) & 3644 \\
        $\mu_1$ & 0.1 & 0.117 & (-0.330,  0.521) & 24502 & 0.124 & (-0.270,  0.497) & 27939 \\
        $\mu_2$ & -0.3 & -0.280 & (-0.550,  0.006) & 30000 & -0.289 & (-0.588,  0.043) & 22964 \\
        $\sigma_{1}^2$ & 0.5 & 0.553 & ( 0.295,  0.915) & 1538 & 0.495 & ( 0.307,  0.751) & 2314 \\
        $\sigma_{2}^2$ & 0.5 & 0.554 & ( 0.150,  1.226) & 496 & 0.512 & ( 0.269,  0.865) & 1329 \\
        $\nu^2$ & 1 & 0.979 & ( 0.875,  1.093) & 5629 & 1.003 & ( 0.935,  1.075) & 13198 \\
        \hline
    \end{tabular}
    \caption{The posterior means, 95\% credible intervals (CI), and effective sample size (ESS) in the case of known basis functions for $\phi _{1} = \phi _{2} = 0.95$. The estimated values of $\sigma_{1}^2$ and $\sigma_{2}^2$ are multiplied by 100, and that of $\nu^2$ by 10000. }
    \label{tab:sim_95_mcmc}
\end{table}

\subsection{The estimation of the Lorenz curves under various models}

An in-depth comparison is made via the estimation of functional value $f_t(x)$ for $x = 0.2, 0.4, 0.6, 0.8$. 
Table~\ref{tab:sim_95_Lorenz} presents the RMSE of posterior means, AL and CP of $95\%$ credible intervals of $f_t(x)$, averaged over $t = 1,\ldots, T$. 
The FSSM with the oracle basis functions attain the smallest RMSEs, as expected, since it
uses the same basis functions as in the data generating process. 
With the misspecified Pareto basis functions, the RMSEs are almost as small as those of the oracle model. This result implies the robustness and flexibility of the FSSMs, although the choice of sufficiently large $L$ and appropriate basis functions are still important. The RMSEs of the ARGPs and DLMs are much higher than those of the FSSMs, showing the limitation of their model flexibility in functional estimation under the shape constraints. Note that the difference between the vanilla and projected ARGPs is almost negligible. In fact, most of the functional values sampled from the vanilla ARGP satisfied the shape constraints.

\begin{table}[htb!]
\centering
\begin{tabular}{cccccccccccccc}
\hline
&& \multicolumn{2}{c}{FSSM} & \multicolumn{2}{c}{ARGP}& \multicolumn{2}{c}{DLM}\\
&  & oracle & misspecified & vanilla & projected & IW & SV\\
\hline
 & $f_t(0.2)$ & 0.209 & 0.479 & 0.691 & 0.691 & 0.390 & 0.366 \\
RMSE & $f_t(0.4)$ & 0.348 & 0.362 & 0.575 & 0.575 & 0.437 & 0.529 \\
($K = 4$) & $f_t(0.6)$ & 0.370 & 0.449 & 0.583 & 0.583 & 0.509 & 0.600 \\
 & $f_t(0.8)$ & 0.406 & 0.495 & 1.077 & 1.077 & 0.764 & 0.604 \\
 \hline
 & $f_t(0.2)$ & 0.172 & 0.382 & 0.545 & 0.546 & 0.352 & 0.346 \\
RMSE & $f_t(0.4)$ & 0.288 & 0.329 & 0.506 & 0.506 & 0.414 & 0.561 \\
($K = 9$) & $f_t(0.6)$ & 0.312 & 0.376 & 0.476 & 0.475 & 0.431 & 0.604 \\
 & $f_t(0.8)$ & 0.333 & 0.401 & 0.618 & 0.628 & 0.556 & 0.596 \\
 \hline
 & $f_t(0.2)$ & 0.965 & 0.660 & 0.995 & 0.995 & 0.995 & 0.895 \\
CP & $f_t(0.4)$ & 0.960 & 0.950 & 1.000 & 1.000 & 0.995 & 0.820 \\
($K = 4$) & $f_t(0.6)$ & 0.960 & 0.895 & 0.995 & 0.995 & 0.980 & 0.765 \\
 & $f_t(0.8)$ & 0.940 & 0.895 & 0.990 & 0.990 & 0.985 & 0.815 \\
\hline
 & $f_t(0.2)$ & 0.960 & 0.655 & 0.985 & 0.980 & 1.000 & 0.920 \\
CP & $f_t(0.4)$ & 0.945 & 0.920 & 0.995 & 0.995 & 1.000 & 0.825 \\
($K = 9$) & $f_t(0.6)$ & 0.930 & 0.905 & 1.000 & 1.000 & 1.000 & 0.800 \\
 & $f_t(0.8)$ & 0.930 & 0.885 & 0.990 & 0.985 & 0.995 & 0.790 \\
 \hline
 & $f_t(0.2)$ & 0.008 & 0.010 & 0.043 & 0.043 & 0.020 & 0.013 \\
AL & $f_t(0.4)$ & 0.014 & 0.015 & 0.040 & 0.040 & 0.021 & 0.014 \\
($K = 4$) & $f_t(0.6)$ & 0.015 & 0.015 & 0.040 & 0.040 & 0.026 & 0.015 \\
 & $f_t(0.8)$ & 0.016 & 0.016 & 0.043 & 0.043 & 0.035 & 0.016 \\
 \hline
 & $f_t(0.2)$ & 0.007 & 0.008 & 0.030 & 0.030 & 0.019 & 0.013 \\
AL & $f_t(0.4)$ & 0.011 & 0.012 & 0.029 & 0.029 & 0.022 & 0.015 \\
($K = 9$) & $f_t(0.6)$ & 0.012 & 0.012 & 0.029 & 0.029 & 0.024 & 0.015 \\
 & $f_t(0.8)$ & 0.012 & 0.013 & 0.030 & 0.030 & 0.031 & 0.016 \\
\hline
\end{tabular}
\caption{The root mean squared errors (RMSE) multiplied by 100, coverage probability (CP) and average length (AL) of 95\% credible intervals of $f_t(x)$ for $x = 0.2, 0.4, 0.6, 0.8$, averaged over $t = 1,\ldots, T$.}
\label{tab:sim_95_Lorenz}
\end{table}

\begin{figure}[htb!]
\centering
\includegraphics[width=\linewidth]{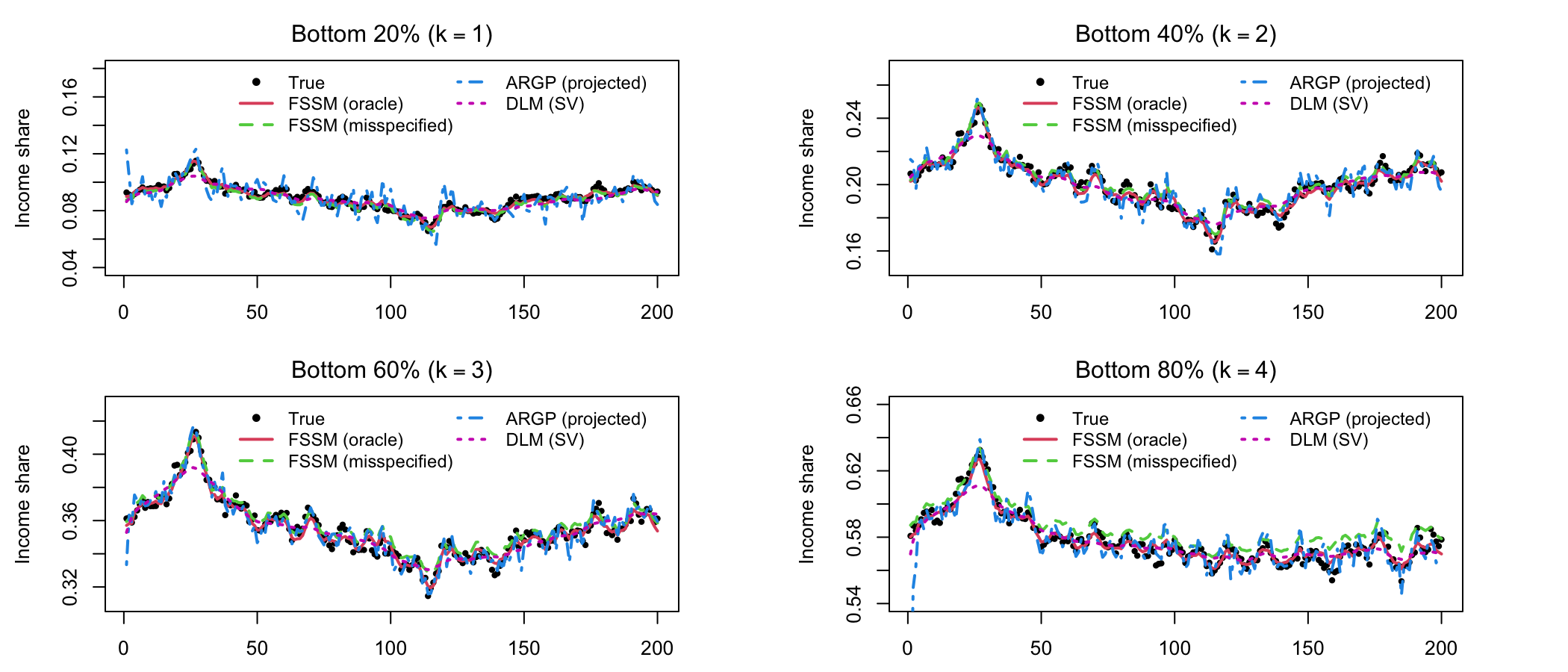}
\caption{Time series of the true and posterior means of the income shares for the bottom 20, 40, 60, and 80\% obtained by FSSM using ``oracle" and ``misspecified" basis functions, ARGP (projected) and DLM with SV, applied to the simulated data with $K = 4$.}
\label{fig:sim_Lorenz_K=4}
\end{figure}

\begin{figure}[htb!]
\centering
\includegraphics[width=\linewidth]{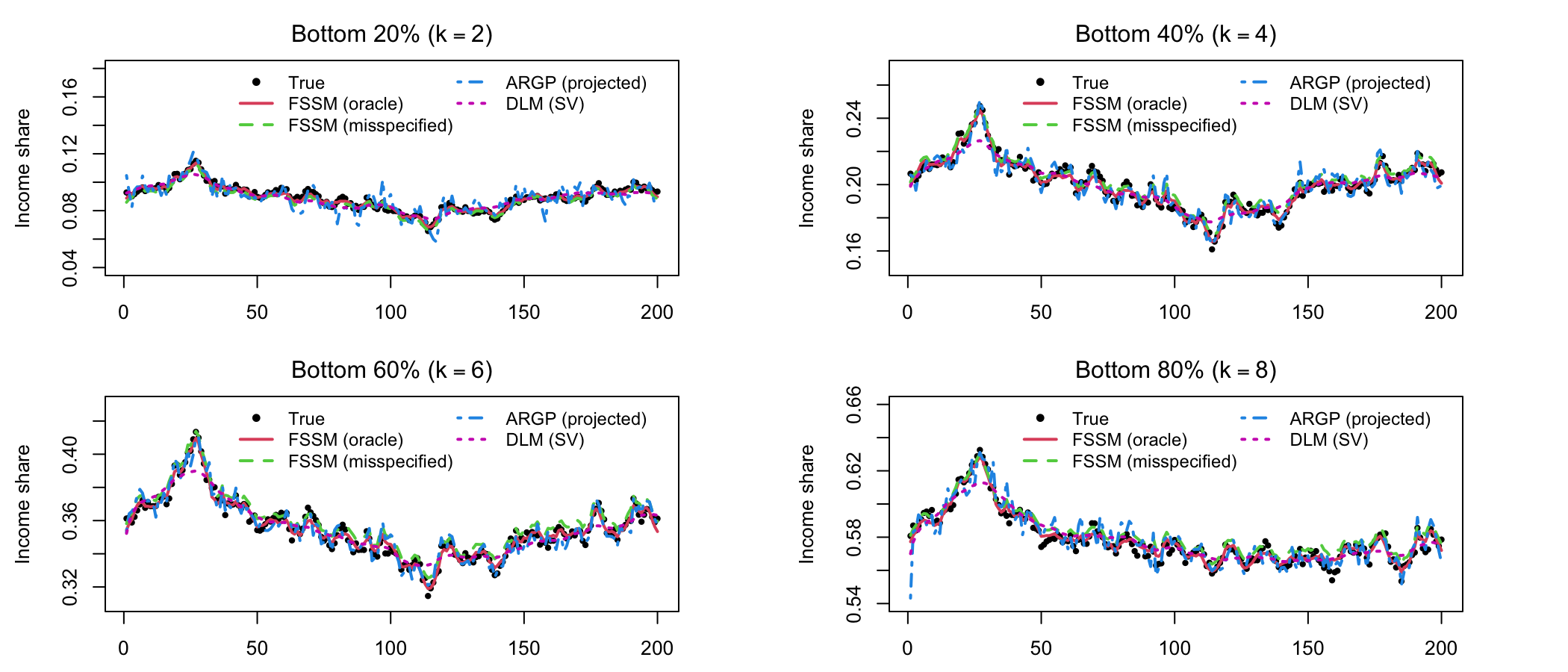}
\caption{Time series of the true and posterior means of the income shares for the bottom 20, 40, 60, and 80\% obtained by FSSM using ``oracle" and ``misspecified" basis functions, ARGP (projected) and DLM with SV, applied to the simulated data with $K = 9$.}
\label{fig:sim_Lorenz_K=9}
\end{figure}

Figures~\ref{fig:sim_Lorenz_K=4} and \ref{fig:sim_Lorenz_K=9} are the time series plots of the posterior means of $f_t(x)$ under the FSSMs, ARGPs and DLMs. 
It is observed that in both cases ($K=4$ and $9$), the proposed FSSM precisely estimates the true values of $f_t(x)$, and the difference between ``oracle" and ``misspecified" basis functions is limited. 
On the other hand, the estimates of ARGPs are highly variable than the true values. The DLMs provide over-smoothed estimates; this explains the large bias in posterior predictive analysis discussed in the main text. The increase of available information improves the model fit. For example, it is visually clear that the estimates of $f_t(0.2)$ under the APRG are less volatile when $K=9$.

\section{Detailed MCMC results in Japan Income Survey Data}
We provide posterior summary (posterior means and $95\%$ credible intervals) of the unknown model parameters in FSSM (with Basis Set 1) in Table~\ref{tab:real_mcmc}. 
It is confirmed that ESS are sufficiently large for all the parameters.
The estimates of $\phi _{\ell}$ are close to unity, especially for $\ell = 1,2$, implying high autocorrelations of dynamic convex weights.

\begin{table}[htb!]
    \centering
    \begin{tabular}{c|ccc}
        \hline
        Parameter & Mean & 95\% CI & ESS \\
        \hline
        $\phi_{11}$ & 0.942 & ( 0.844,  0.996) & 585 \\
        $\phi_{22}$ & 0.959 & ( 0.911,  0.995) & 1775 \\
        $\phi_{33}$ & 0.805 & ( 0.500,  0.986) & 470 \\
        $\phi_{44}$ & 0.821 & ( 0.598,  0.963) & 408 \\
        $\mu_1$ & 1.384 & ( 1.202,  1.525) & 673 \\
        $\mu_2$ & -0.027 & (-0.259,  0.175) & 2939 \\
        $\mu_3$ & -0.059 & (-0.511,  0.028) & 281 \\
        $\mu_4$ & -0.045 & (-0.165,  0.002) & 270 \\
        $\sigma_{11}^2$ & 0.061 & ( 0.022,  0.117) & 617 \\
        $\sigma_{22}^2$ & 0.161 & ( 0.093,  0.239) & 561 \\
        $\sigma_{33}^2$ & 0.178 & ( 0.024,  0.597) & 289 \\
        $\sigma_{44}^2$ & 0.059 & ( 0.017,  0.124) & 339 \\
        $\nu^2$ & 0.023 & ( 0.020,  0.026) & 1235 \\
        \hline
    \end{tabular}
    \caption{The posterior means, 95\% credible intervals (CI) of parameters of FSSM (with Basis Set 1), applied to Japanese income survey data. 
    The estimated values of $\sigma_{\ell}^2$ for $\ell=1, \dots, 4$ are multiplied by 100, and that of $\nu^2$ by 10000. }
    \label{tab:real_mcmc}
\end{table}

\begin{figure}[htbp]
\centering
\includegraphics[width=0.8\linewidth]{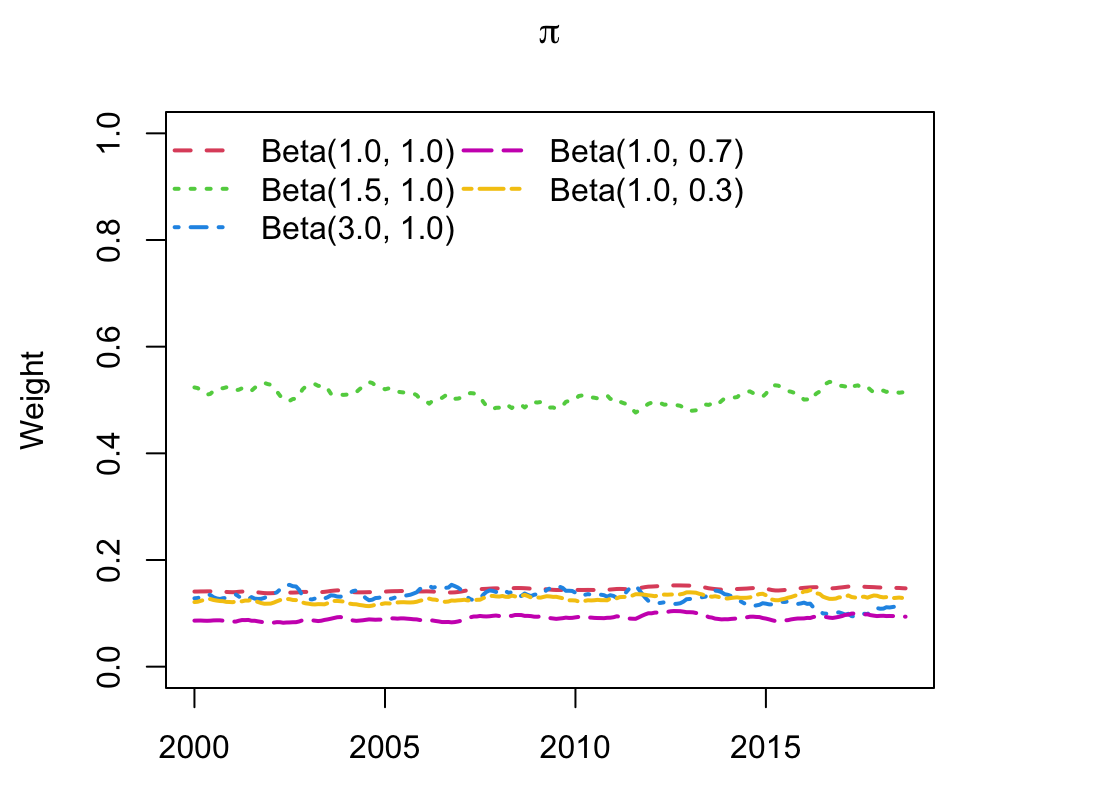}
\caption{The posterior means of $\pi_{t1},\ldots, \pi_{t5}$ of FSSM.}
\label{fig:real_rough_5_pi}
\end{figure}

Figure~\ref{fig:real_rough_5_pi} is the time series plots of the posterior means of $\pi_{t1}, \ldots, \pi_{t5}$. 
The dynamics of the estimated weights might look subtle visually, but surely contribute to the volatile behaviors of the Lorenz curves and Gini coefficients in Figures~\ref{fig:real_Gini1} and \ref{fig:real_Gini2}. 
The largest weight is placed on the basis function of Beta(1.5, 1.0), which is the closest one to the observed functional values on $(x_1,\dots x_4)$.

\end{document}